\documentclass[fleqn,usenatbib]{mnras}

\usepackage[T1]{fontenc}
\usepackage{ae,aecompl}

\usepackage{graphicx}	% Including figure files
\usepackage{amsmath}	% Advanced maths commands
\usepackage{amssymb}	% Extra maths symbols
\usepackage{breakurl}

\title[An eLIMA model for CAL~83]{An eLIMA model for the 67~s X-ray periodicity in CAL~83}

\author[A. Odendaal and P. J. Meintjes]{
A. Odendaal$^{1}$\thanks{E-mail: WinkA@ufs.ac.za}
and P. J. Meintjes$^{1}$
\\
$^{1}$Department of Physics, University of the Free State, P.O. Box 339, Bloemfontein, 9300, South Africa
}

\date{Accepted XXX. Received YYY; in original form ZZZ}

\pubyear{2016}

\begin{document}
\label{firstpage}
\pagerange{\pageref{firstpage}--\pageref{lastpage}}
\maketitle

% Abstract of the paper
\begin{abstract}
Supersoft X-ray sources (SSSs) are characterized by their low effective temperatures and high X-ray luminosities. The soft X-ray emission can be explained by hydrogen nuclear burning on the surface of a white dwarf (WD) accreting at an extremely high rate. A peculiar $\sim$67~s periodicity ($P_{67}$) was previously discovered in the \textit{XMM-Newton} light curves of the SSS CAL~83. $P_{67}$ was detected in X-ray light curves spanning $\sim$9~years, but exhibits variability of several seconds on time-scales as short as a few hours, and its properties are remarkably similar to those of dwarf nova oscillations (DNOs). DNOs are short time-scale modulations ($\lesssim$1~min) often observed in dwarf novae during outburst. DNOs are explained by the well established low-inertia magnetic accretor (LIMA) model. In 
this paper, we  show
that $P_{67}$ and its associated period variability can be satisfactorily explained by an application of the LIMA model to the more `extreme' environment in a SSS (eLIMA), contrary to another recent study attempting to explain  $P_{67}$ and its associated variability in terms of non-radial g-mode oscillations in the extended envelope of the rapidly accreting white dwarf in CAL~83. In the eLIMA model, $P_{67}$ originates in an equatorial belt in the WD envelope at the boundary with the inner accretion disc, with the belt weakly coupled to the WD core by a $\sim$10$^5$~G magnetic field. 

New optical light curves obtained with the Sutherland High-speed Optical Camera (SHOC) are also presented, exhibiting quasi-periodic modulations on time-scales of $\sim$1000~s, compatible with the eLIMA framework.
\end{abstract}

% Select between one and six entries from the list of approved keywords.
% Don't make up new ones.

%%%%%%%%%%%%%%%%%%%%%%%%%%%%%%%%%%%%%%%%%%%%%%%%%%

%%%%%%%%%%%%%%%%% BODY OF PAPER %%%%%%%%%%%%%%%%%%

\section{Introduction}

Supersoft X-ray sources (SSSs) were established as a quite unique class of objects after observations by the \textit{Einstein Observatory} \citep{Long_etal1981,SewardMitchell1981} and \textit{ROSAT} \citep{Trumper_etal1991}. Their defining characteristics are their low X-ray temperatures of $kT_{\rm eff}\sim20$--100~eV, and their extreme soft X-ray luminosities, from $\sim$10$^{36}{\rm~erg~s}^{-1}$ up to $\sim$10$^{38}{\rm~erg~s}^{-1}$ (e.g.~\citealp{KahabkaVandenHeuvel2006}). The vast majority of observed SSSs are binary systems. \cite{VandenHeuvel_etal1992} showed that their observational properties can be explained by the nuclear burning of accreted hydrogen on the surface of a white dwarf (WD).

CAL~83 was one of the first SSSs to be discovered, and with its relatively short orbital period of $1.047529\pm0.000001$~d \citep{Rajoelimanana_etal2013} it forms part of the so-called `close binary supersoft source' (CBSS) subclass. \cite{Lanz_etal2005a} fitted WD atmosphere models to \textit{Chandra} and \textit{XMM-Newton} X-ray spectra, and the best-fitting WD mass was $M_1\sim1.3{\rm~M}_\odot$, with a luminosity of $L\sim3.4\times10^{37}{\rm~erg~s}^{-1}$. \cite{OdendaalMeintjes2015b} provided a general review of the properties of CAL~83, while \cite{OdendaalMeintjes2015a} considered the source variability.

CAL~83 exhibits long-term quasi-periodic modulations in the optical ($P\sim450$~d), cycling between an optical low state and an optical high state (\citealp{Rajoelimanana_etal2013} and references therein). Several X-ray off-states have been observed during optical high states, while the X-ray high states were observed during optical low states. This long-term anti-correlation between X-ray and optical flux are also observed in another LMC SSS, namely RX~J0513.9--6951, although the latter has a shorter cycle period of $\sim$168~d. Several authors have discussed different aspects of a so-called `limit cycle' model that can be employed to explain this behaviour in both sources.
In such a model, the X-ray high, optical low state is associated with the WD photosphere being in a contracted state (higher effective temperature), while the X-ray low, optical high state corresponds to an expanded WD photosphere (lower effective temperature) due to an enhanced accretion rate \citep{Southwell_etal1996a, Reinsch_etal2000, HachisuKato2003b}.

CBSSs are closely related to another class in the WD binary population, namely the cataclysmic variables (CVs). In CVs, the primary WD is more massive than the donor, i.e.~the mass ratio $q=M_2/M_1$ is smaller than 1. However, it has been shown that an accretion rate of the order $\dot{m}_{\rm acc}\sim10^{-7}{\rm~M}_\odot{\rm~yr}^{-1}$ is required to drive persistent surface nuclear burning in SSSs, which can only be sustained (in the case of Roche lobe overflow) if the mass of the donor is comparable to or greater than the WD mass ($q\gtrsim1$). The main characteristics of SSSs that distinguish them from CVs are therefore the inverted mass ratios of the former, their high $\dot{m}_{\rm acc}$ and the extreme luminosities derived from surface nuclear burning.

To provide some context for the high $\dot{m}_{\rm acc}$ in SSSs, it is useful to keep in mind the range of $\dot{m}_{\rm acc}$ that is typical of different subclasses in the related class of CVs:  from $\sim$(2--50)$\times10^{-11}{\rm~M}_\odot{\rm~yr}^{-1}$ in dwarf novae during quiescence, to $\sim$(3--10)$\times10^{-9}{\rm~M}_\odot{\rm~yr}^{-1}$ in nova remnants, novalikes and dwarf novae during outburst (e.g.~\citealp[pp.~64--66]{Warner1995a}).

It is well known that CVs exhibit (often quite drastic) variability on various time-scales, including quasi-periodic modulations with periods ranging from a few seconds to a few thousand seconds. These modulations have been well studied in CVs, but due to the great similarity between CVs and supersoft X-ray binaries, one can expect similar quasi-periodic modulations from the latter class.

\cite{Odendaal_etal2014a} reported the discovery of a $\sim$67~s X-ray periodicity (hereafter $P_{67}$) in archival \textit{XMM-Newton} data of CAL~83. The peculiar variability observed in this period precluded a straightforward interpretation as the WD rotation period. The favoured interpretation of \cite{Odendaal_etal2014a} was that the period may originate in an extended WD envelope of which the rotation is not quite synchronized with the rotation of the WD itself, causing the observed drift in the period. Since the discovery, a detailed comparison of $P_{67}$ with the numerous different types of periodicities observed in CVs has shown that it is remarkably similar to dwarf nova oscillations (DNOs), which are variable, short time-scale  oscillations observed in dwarf novae during outburst, as well as in other high-$\dot{m}_{\rm acc}$ CVs. This provides a suitable theoretical framework to investigate the possible correlation between $P_{67}$ and its associated variability in CAL83, and DNOs, which differs from the recent study 
(\citealp{Ness_etal2015}) which attempted to explain this phenomenon in terms of oscillating non-radial g-mode oscillations in the extended envelope of the rapidly rotating white dwarf. 

The most widely accepted model for DNOs involves anisotropic emission from a region close to the surface of a WD of which the magnetic field is too weak to enforce rigid co-rotation of the core and exterior regions, but strong enough to significantly influence accretion close to the WD surface. This is known as the low-inertia magnetic accretor (LIMA) model \citep{WarnerWoudt2002}. In this paper, the application of a similar LIMA model to the observed $P_{67}$ in CAL~83 is discussed. Although the same underlying principles of this model as applied to dwarf novae are still applicable, one must keep in mind that the conditions on the surface of the WD in SSSs are probably more extreme, with the high $\dot{m}_{\rm acc}$ and the associated extended envelope of accreted material with hydrogen burning at the base of the envelope. In this paper, we therefore refer to this application of the LIMA model in the `extreme' accretion limit by the acronym `eLIMA'.

In \S\ref{sec:CV_quasi-periodic_modulations}, a brief summary of the different short time-scale quasi-periodic modulations observed in CVs is given, while \S\ref{sec:LIMA_DNe} provides an overview of those aspects of the LIMA model for DNOs that are the most relevant for the current discussion. The properties of $P_{67}$ in CAL~83 are reviewed in \S\ref{sec:P67_CAL83}, where the analysis of \cite{Odendaal_etal2014a} has been extended to take into account the effect of the optical blocking filters of \textit{XMM-Newton}, and to better illustrate the dynamical nature of $P_{67}$.  The proposed properties of the eLIMA model in CAL~83 are discussed in \S\ref{sec:eLIMA_CAL83}. 
New optical light curves of CAL~83 have been obtained with the Sutherland High-speed Optical Camera (SHOC) on the 1.9-m Telescope at the South African Astronomical Observatory (SAAO), and the modulations observed in these light curves support the proposed eLIMA model. These are presented in \S\ref{sec:SHOC}, followed by the conclusion in \S\ref{sec:Conclusions}.

\section{Quasi-periodic modulations in cataclysmic variables}
\label{sec:CV_quasi-periodic_modulations}

Many CVs have been observed to exhibit quasi-periodic variability. The time-scales of these oscillations are typically between a few seconds and a few thousand seconds (see e.g.~the review papers of \citealp{warner2004} and \citealp{WarnerWoudt2008}). The first type of oscillation is known as `dwarf nova oscillations' (DNOs), as they were first discovered in dwarf novae (DNe) during outburst. However, DNOs have also been observed in other CVs with high accretion rates, but never in intermediate polars (IPs). IPs are characterized by magnetic fields of the order of $10^7{\rm~G}$.

The typical periods associated with DNOs are in the $\sim$5--40~s range. They have primarily been observed in the optical, with typical oscillation amplitudes of $<$1~per~cent, and the pulse profile is highly sinusoidal in shape. However, DNOs have also been discovered in the ultraviolet and soft X-rays with much larger amplitudes, even up to 100~per~cent in soft X-rays. The rate of change of a quasi-periodic period $P$ is often described by the quality factor
\begin{equation}
 Q\equiv \left| \frac{dP}{dt} \right|^{-1}~,
\end{equation}
which will have a small value if there are rapid changes in $P$, while a highly coherent period will have a large $Q$-value. The value of $Q$ for DNOs is usually in the $10^3$--$10^7$ range, while e.g.~the coherent WD rotation period in the IP DQ~Her has $Q\sim10^{12}$.

DNO periods exhibit variability of both a continuous and a discontinuous nature. The continuous variability obeys a period-luminosity relation, with the shortest period corresponding to the highest luminosity, i.e.~the highest $\dot{m}_{\rm acc}$ during the DN outburst, and the period increases as the post-outburst luminosity decreases. However, this relation is actually not one-to-one, but rather has a `banana loop' nature (see \citealp{Patterson1981} for more details). Discontinuous jumps of $\sim$0.01~per~cent are also sometimes observed in the period.

Quasi-periodic modulations belonging to the second class are known as `longer period DNOs' (lpDNOs) \citep{Warner_etal2003}. They have periods typically $\sim$4 times longer than DNOs, and slightly larger amplitudes, without the strong period-luminosity relationship of DNOs. These lpDNOs are also only observed in CVs with high accretion rates.

The third type is simply called `quasi-periodic oscillations' (QPOs), with periods and oscillation amplitudes much larger than those of DNOs. QPOs are also less coherent than DNOs, with $Q$ typically between 5 and 20. QPOs have not only been observed in high $\dot{m}_{\rm acc}$ CVs, but also in DNe during quiescence. In cases where QPOs do occur simultaneously with DNOs, the relation $P_{\rm QPO}\sim15 P_{\rm DNO}$ has been observed, and then the term `DNO-related QPO' is often used \citep{Warner_etal2003}.

Another type of QPO also exists, having much longer periods of $\sim$1000--3000~s. Many CVs with such a QPO are suspected to be IPs, where the coherent WD spin period is not observed directly, possibly due to a high accretion rate, but reprocessed by a QPO source with varying period. This is supported by these QPOs having periods comparable to that of the WD rotation periods in canonical intermediate polars, i.e.~$\sim$15~min. However, in cases where QPOs are observed simultaneously with DNOs, the system can not be an intermediate polar (see below). In these systems, the longer period QPOs may represent the Keplerian periods of blobs at the outer edge of an accretion disc. Another source of QPOs may be a modulation in mass transfer caused by non-radial oscillations of the secondary.

\section{The existing LIMA model for dwarf nova oscillations}
\label{sec:LIMA_DNe}

It has been proposed that DNOs originate in systems where the accreting WD has a weak magnetic field, thus representing an extension of the intermediate polar class to lower magnetic field values. In this regard, it is important to consider the findings of \cite{Katz1975}. He considered the rotation period $P_*$ of the WD in DQ~Her, and argued that the high degree of coherency of the observed period must be a result of rigid body rotation of the WD, with a magnetic field as the mechanism coupling the stable clock of the rotating core to the outer environment.

The magnetic field required to transmit this angular acceleration from the WD surface regions to the core can be estimated by equating the Maxwell and mechanical stresses, i.e.
\begin{equation}
 \frac{B_r B_\phi}{4\pi} \approx  \rho_{\rm wd}~\dot{\Omega}_*R_1^2~,
 \label{eq:Katz_B}
\end{equation}
where $B_r$ and $B_\phi$ represent the radial and azimuthal components of the internal WD magnetic field respectively. A typical WD radius of $R_1\sim10^9{\rm~cm}$ can be used in the equation above, and assuming that a significant portion of the deep interior of the WD is still in fluid form and not yet crystalline, a density of $\rho_{\rm wd}\sim10^6{\rm~g~cm}^{-3}$ can be used. Adopting these parameters yields $B_r B_\phi\sim10^{10}{\rm~G}^2$, and approximating the two components as being equal, one obtains $B_r\sim B_\phi\sim10^5{\rm~G}$ \citep{Katz1975}. This provides an approximate lower limit for the magnetic field to prevent differential rotation of the outer regions.

For weaker magnetic fields, the accretion torque will not be coupled rigidly to the interior, allowing the formation of a belt structure in the equatorial region on the surface on the WD that is rotating rapidly at the local Keplerian velocity, in other words faster than the central WD core. Actually, due to its inertia, this `equatorial belt' would most likely have a period slightly longer than the local Keplerian period \citep{Warner1995b}. In fact, evidence supporting the existence of a hot, rapidly rotating belt around a WD in relatively slow rotation has been found in HST spectra of the dwarf novae VW~Hyi, U~Gem, AL~Com and RU~Peg during and after outburst \citep{Sion_etal1996,GansickeBeuermann1996,Cheng_etal1997,Szkody_etal1998,SionUrban2002}. This concept was initially suggested by \cite{Paczynski1978}, where-after \cite{Warner1995b} and \cite{WarnerWoudt2002} developed it into a low-inertia magnetic accretor (LIMA) model to explain the DNO phenomenon.

Within the LIMA model, the total accretion luminosity is used to estimate the total mass of the material contained in the equatorial belt as $<$10$^{-10}{\rm~M}_\odot$. Therefore, the inertia of the belt is much lower than the inertia of the whole WD, and the belt can readily be tugged around due to magnetic coupling with the inner disc. Then, the DNO period-luminosity relation can be explained as follows:  As $\dot{m}_{\rm acc}$ increases, the WD magnetosphere is compressed and the inner disc boundary moves further inward, yielding a shorter Keplerian (or quasi-Keplerian) period for the belt (spin-up). As $\dot{m}_{\rm acc}$ decreases during the final stages of outburst, the inner disc boundary moves outward rapidly. It is expected that the belt would not be able to also slow down rapidly enough to remain near equilibrium, and some of the material in the belt is centrifuged to larger radii, removing angular momentum from the belt (spin-down).

Although the spin-up and spin-down mechanisms above could explain the continuous changes in DNO periods, it can not explain the discontinuous `jumps' that are also observed. The latter are most likely caused by magnetic reconnection events, through which the system adjusts to seek the accretion configuration with the lowest energy. Even though the magnetic field is relatively weak, it will still have significant interaction with the accreting material. Magnetic field lines threading the slipping equatorial belt will be wound up, until reconnection starts occurring after differential rotation of $\sim$2$\pi$. The whole equatorial belt is not necessarily associated with a single period. Latitudinal variations in rotational period may be present in the belt, with longer periods at higher latitudes (analogous to the latitudinal differential rotation observed in the Sun, e.g.~\citealp[p.~364]{CarrollOstlie2007}), approaching the rotation period of the underlying primary. 
Magnetohydrodynamic (MHD) turbulence caused by reconnection may be associated with so-called `accretion curtains' where gas can potentially be fed through accretion arcs at different latitudes.

One can expect that many such accretion arcs would be involved simultaneously, and if these arcs are spread out over different latitudes in a quite regular manner, then a relatively coherent period may not be observable at all. On the other hand, the coincidental domination of one or two arcs may yield an observable DNO. A reconnection event would cause a reconfiguration of one or more accretion arcs, and the rapid transference of the accretion stream from one arc to the other would be able to explain the observed jumps in the observed DNO period.

In several systems, two DNOs with slightly different periods have been observed, together with a DNO-related QPO, and it has been found that the slightly longer DNO corresponds to the beat period between the shorter DNO and the QPO. In such a case, the shorter DNO could be associated with the equatorial belt as explained above, and the QPO with a rotating structure (like a thickening in the accretion disc, e.g.~at the point of impact of the accretion stream) revolving around the primary in a prograde direction. The beat period is then caused by the reflection/reprocessing of the shorter DNO beam by the QPO structure.

The lpDNOs are also thought to be associated with such systems, but are found to have periods equal to approximately one half of the spin period of the WD itself. \cite{warner2004} suggested that this connects the lpDNOs directly to the WD spin period, with the factor of 2 being due to magnetically channelled two-pole accretion onto the WD. Because $P_{\rm lpDNO} \sim 4 P_{\rm DNO}$, this implies that the equatorial belt at the inner disc is rotating at an angular velocity 8 times larger than that of the WD itself.

\section{The nature of the 67~\lowercase{s} periodicity in CAL~83}
\label{sec:P67_CAL83}

There are 23 observations of CAL~83 in the \textit{XMM-Newton} Science Archive. One of these observations was obtained in 2000, five in 2007, sixteen in 2008 and one in 2009 (see Table~\ref{tab:CAL83_XMM-Newton_all_obs}). During all the observations, data were obtained with all the X-ray detectors on board, i.e.~the three EPIC detectors (MOS1, MOS2 and pn) \citep{struder_etal2001,turner_etal2001} and the Reflection Grating Spectrometers (RGS). All the observations, except the first, also included Optical Monitor (OM) data, but these OM data do not form part of the current discussion.

\begin{table*}
\centering
\caption{Archival \textit{XMM-Newton} observations of CAL~83. The last 3 columns provide information regarding the $\sim$67~s periodicity, for those observations where a periodogram peak at this position was detected.}
\label{tab:CAL83_XMM-Newton_all_obs}
\begin{tabular}{@{}cl@{~~}r@{~~~}c@{~~~}c@{~~~}c@{~~~}c@{~~}c@{~~~}c@{}}
\hline
Observation 	&\multicolumn{2}{c}{Start date \&}	&Duration	& Mean EPIC 		&X-ray	&Period	&Significance	&Modulation\\
ID 		&\multicolumn{2}{c}{time (UT)}		&(s)		& pn CR$^\text{a}$ 	&state	&(s)	&(per cent)	&semi-amplitude (per cent)\\
\hline
0123510101A	&2000~Apr~23  &07:34:01	&45021	&$6.471\pm0.021$&Bright	&$68.69\pm0.24$ &29.94	&2.5\\
0123510101B	&&&&&&$66.86\pm0.22$ &59.47	&2.8\\
0123510101C	&&&&&&$66.63\pm0.22$ &98.16	&5.0\\
0500860201	&2007~May~13  &22:03:32	&11949	&$7.324\pm0.036$ &Bright	&$65.45\pm0.19$ &81.69	&3.1\\
0500860301	&2007~Jul~06  &23:31:52	&10920	&$6.764\pm0.036$ &Bright	&$67.55\pm0.22$ &99.02	&4.0\\
0500860401	&2007~Aug~21  &15:11:21	&7915	&$5.208\pm0.037$ &Bright	&$66.55\pm0.30$ &23.25	&3.9\\
0500860501	&2007~Oct~05  &23:49:21	&17615	&$5.034\pm0.028$ &Bright	&$70.55\pm0.20$ &51.74	&4.0\\
0500860601	&2007~Nov~24  &21:10:14	&23173	&$6.445\pm0.026$ &Bright	&$67.67\pm0.11$ &99.99	&3.7\\
0500860701	&2008~Jan~16  &13:24:33	&10914	&				&Off/Low	&&&\\
0500860801	&2008~Mar~10  &10:14:38	&6916	&				&Off/Low	&&&\\
0506530201	&2008~Mar~20  &00:33:22	&7715	&$0.619\pm0.013$ &Faint		&&&\\
0506530301	&2008~Apr~03  &18:40:55	&14918	&				&Off/Low	&&&\\
0506530401	&2008~Apr~11  &06:02:24	&5914	&				&Off/Low	&&&\\
0506530501	&2008~Apr~16  &12:32:01	&10873	&$1.680\pm0.025$ &Faint		&$66.76\pm0.49$ &56.03	&8.6\\
0506530601	&2008~Apr~17  &13:40:13	&11217	&$1.426\pm0.015$ &Faint		&&&\\
0506530801	&2008~Apr~19  &06:43:35	&5916	&$0.350\pm0.012$ &Faint		&&&\\
0506530901	&2008~Apr~20  &22:38:43	&11617	&$0.403\pm0.008$ &Faint		&&&\\
0500860901	&2008~Apr~21  &02:10:24	&12716	&$0.234\pm0.008$ &Faint		&&&\\
0506531001	&2008~Apr~21  &18:47:57	&9076	&$0.507\pm0.012$ &Faint		&&&\\
0506531201	&2008~Apr~23  &11:20:22	&7418	&$0.166\pm0.008$ &Faint		&&&\\
0506531301	&2008~Apr~25  &08:13:32	&9614	&$0.438\pm0.010$ &Faint		&&&\\
0506531401	&2008~Apr~29  &00:40:23	&14117	&$0.249\pm0.006$ &Faint		&&&\\
0506531501	&2008~Aug~12  &14:50:27	&6918	&$7.846\pm0.045$ &Bright	&$66.87\pm0.35$ &$\gg$99.99	&7.6\\
0506531601	&2008~Sep~17  &11:10:21	&6817	&$0.101\pm0.007$ &Faint		&&&\\
0506531701	&2009~May~30  &08:00:48	&46115	&$7.687\pm0.017$ &Bright	&$65.173\pm0.047$ &$\gg$99.99	&2.5\\
\hline
\end{tabular}
$^\text{a}$Converted to `thin filter counts' (see text).
\end{table*}

Four of the observations were performed during an X-ray off-state. The 19 X-ray on-state observations were reduced by following standard data reduction procedures with the \textit{XMM-Newton} Science Analysis System ({\sc sas}) Version 13.0.1\footnote{\url{http://xmm.esac.esa.int/sas/}} \citep{sas_manual2014}. In Table~\ref{tab:CAL83_XMM-Newton_all_obs}, the mean EPIC pn count rate (CR) in the 0.15--1.0~keV band (the `broadband' CR) is provided for each observation (converted to `thin filter counts' as explained below) to enable a comparison of the relative strength of the X-ray emission between different observations. 

During observation 0123510101, three exposures were obtained with each EPIC detector, with thin, medium and thick optical blocking filters respectively. The gaps between the exposures were almost 5~ks, therefore they were analysed separately, and were assigned observation IDs of 0123510101A, 0123510101B and 0123510101C. As there are 19 on-state observations, this subdivision of observation 0123510101 yields a total of 21 on-state datasets to be considered. For the other EPIC datasets, either a thin or a medium filter was used. It must be kept in mind that the choice of optical blocking filter also influences the X-ray CR significantly, especially at soft energies. The correction for the effect of the optical blocking filter was not discussed by \cite{Odendaal_etal2014a}, therefore we provide details here.

To estimate the effect of the optical blocking filter, the \textit{Chandra} proposal planning tool PIMMS v4.7b was used, which provides general functionality for conversions between source flux and CR for a variety of X-ray instruments\footnote{\url{http://cxc.harvard.edu/toolkit/pimms.jsp}}. By using a blackbody approximation, the shape and flux of the CAL~83 X-ray spectrum were provided as input to PIMMS. The tool was then used to estimate the expected CRs in EPIC pn in the 0.15--1.0~keV band, using different optical blocking filters. It was found that, for EPIC pn data, multiplicative factors of 1.2 and 2.6 could be used to convert CRs obtained with the medium and thick filters respectively to an equivalent CR with the thin filter. The same estimates were performed for EPIC MOS, yielding factors of 1.2 and 2.2. 
These factors are model-dependent, and keeping in mind that the long-term X-ray flux changes in CAL~83 are probably related to temperature modulations, these factors may not be valid for a wide range of CRs.

Alternatively, one could assume for the moment that the average X-ray luminosity of CAL~83 was identical during each of the three EPIC segments in observation 0123510101. By calculating the ratios between the mean observed CRs in the 0.15--1.0~keV band for the 3 segments, multiplicative factors of 1.2 and 2.3 for pn, and 1.3 and 2.4  for MOS were obtained, for conversion from medium and thick filter counts respectively to `thin filter counts'. These values compare quite well with the factors predicted by PIMMS. 

Subsequently, the factors derived from observation 0123510101 were applied to all the EPIC observations obtained with medium and thick optical blocking filters, to yield the approximate CR that would have been obtained if a thin filter was used. The benefit of this transformation is that the EPIC CRs for different observations can now be compared directly. Because this was a simple linear transformation of the CRs and errors, it did not affect the subsequent period analysis. Considering the mean EPIC pn CR per observation as given in Table~\ref{tab:CAL83_XMM-Newton_all_obs}, one can broadly divide the \textit{XMM-Newton} observations into two categories (apart from the four X-ray off-states):  an X-ray bright state with ${\rm CR}>5$~counts~s$^{-1}$, and an X-ray faint state with ${\rm CR}<2$~counts~s$^{-1}$. 

The main period analysis method applied in this study was the well-known Lomb-Scargle (LS) periodogram, as implemented in the Starlink 
{\sc period}\footnote{\url{http://www.starlink.rl.ac.uk/docs/sun167.htx/node16.html}} code, version 5.0-2. A detailed LS analysis of the EPIC light curves was performed, experimenting with different detrending methods and time bin sizes. As the EPIC pn detector has the highest effective area and is therefore most suited to timing, the current discussion will mostly focus on the EPIC pn results.

A very strong peak at $\sim$15~mHz was immediately evident in the periodogram of EPIC pn dataset 0506531501, corresponding to a period of $\sim$67~s. This peak was also present at a $>$99.73~per~cent significance level in observations 0500860601 and 0506531701. By simply scrutinizing the rest of the periodograms by eye and comparing the signal strength at 15~mHz with the local noise level, it was concluded that the $\sim$67~s periodicity can also be considered to be `present' in observations 0123510101 (all 3 segments), 0500860201, 0500860301, 0500860401, 0500860501 and 0506530501, while it is `absent' in the rest of the observations. (It is noted that this `absence' is probably a selection bias rather than an intrinsic source property, as one is more likely to detect a short time-scale variation in a light curve with higher counts and therefore better statistics. A detector with a larger effective area might have revealed the same periodicity in the faint observations.)

\begin{figure*}
  \includegraphics{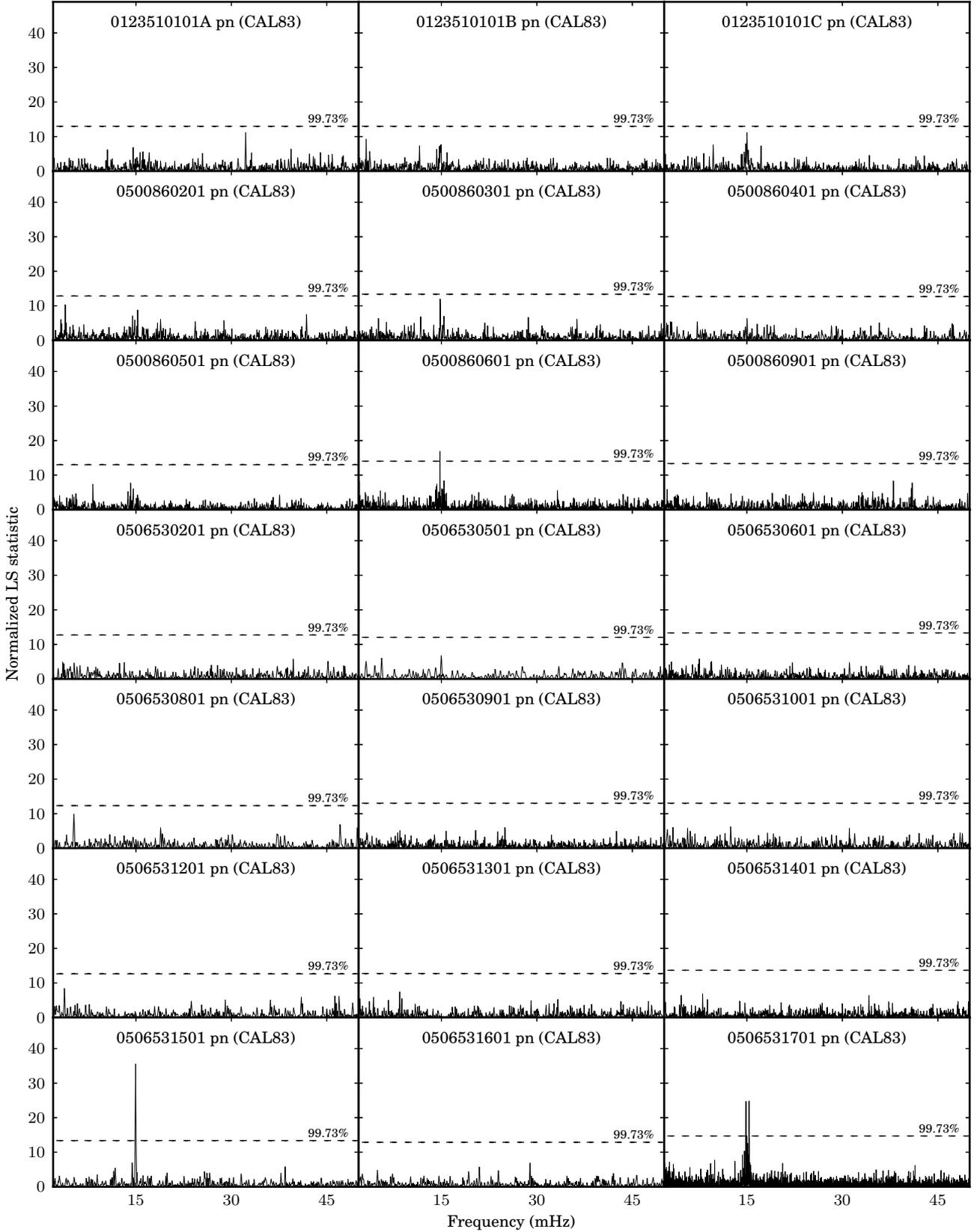}
  \caption{Lomb-Scargle (LS) periodograms of the broadband EPIC pn light curves of CAL~83 in the 2--50~mHz range. The 99.73~per~cent significance level is indicated. A light curve binning of 5~s was used, and the light curves were detrended by subtracting a second-order polynomial fit.}
    \label{fig:pnFIN005aP2H_scrgle}
\end{figure*}

The periodograms of all the EPIC pn on-state light curves of CAL~83 in the 2--50~mHz range are shown in Fig.~\ref{fig:pnFIN005aP2H_scrgle}.  (The 0--100~mHz pn periodograms of some of these observations were presented in figure 1 of \citealp{Odendaal_etal2014a}, together with the MOS periodograms where $P_{67}$ was evident, but here the full set of pn periodograms is provided.) The exact periods associated with the $\sim$15~mHz peak in each of these observations are also listed in Table~\ref{tab:CAL83_XMM-Newton_all_obs}, together with the statistical significance of the peak. The error estimate in the period value was obtained from the simple relation
\begin{equation}
\label{eq:DeltaP}
 \Delta P = \frac{P^2}{2T}~,
\end{equation}
which represents the Fourier resolution of the periodogram, where $P$ is the value of the period itself, and $T$ is the total length of the dataset.

When considering the mean EPIC pn CRs in Table~\ref{tab:CAL83_XMM-Newton_all_obs} together with the results of the LS analysis, one can see that $P_{67}$ was detected in all the X-ray bright observations, with an additional detection in the brightest of the faint state observations: 0506530501, with ${\rm CR}=1.680\pm0.025$. 
The LS periodograms of the much lower signal-to-noise MOS light curves did not exhibit this peak at a $>$99.73~per~cent level, although it did appear slightly above the local noise level in observations 0123510101A, 0123510101B, 0500860301, 0500860401 and especially in 0506531701. This discovery of $P_{67}$ was reported by \cite{Odendaal_etal2014a}, who also showed that this periodicity is inherent to CAL~83 and is not found in the periodograms of the background counts. 

\begin{figure*}
  \includegraphics{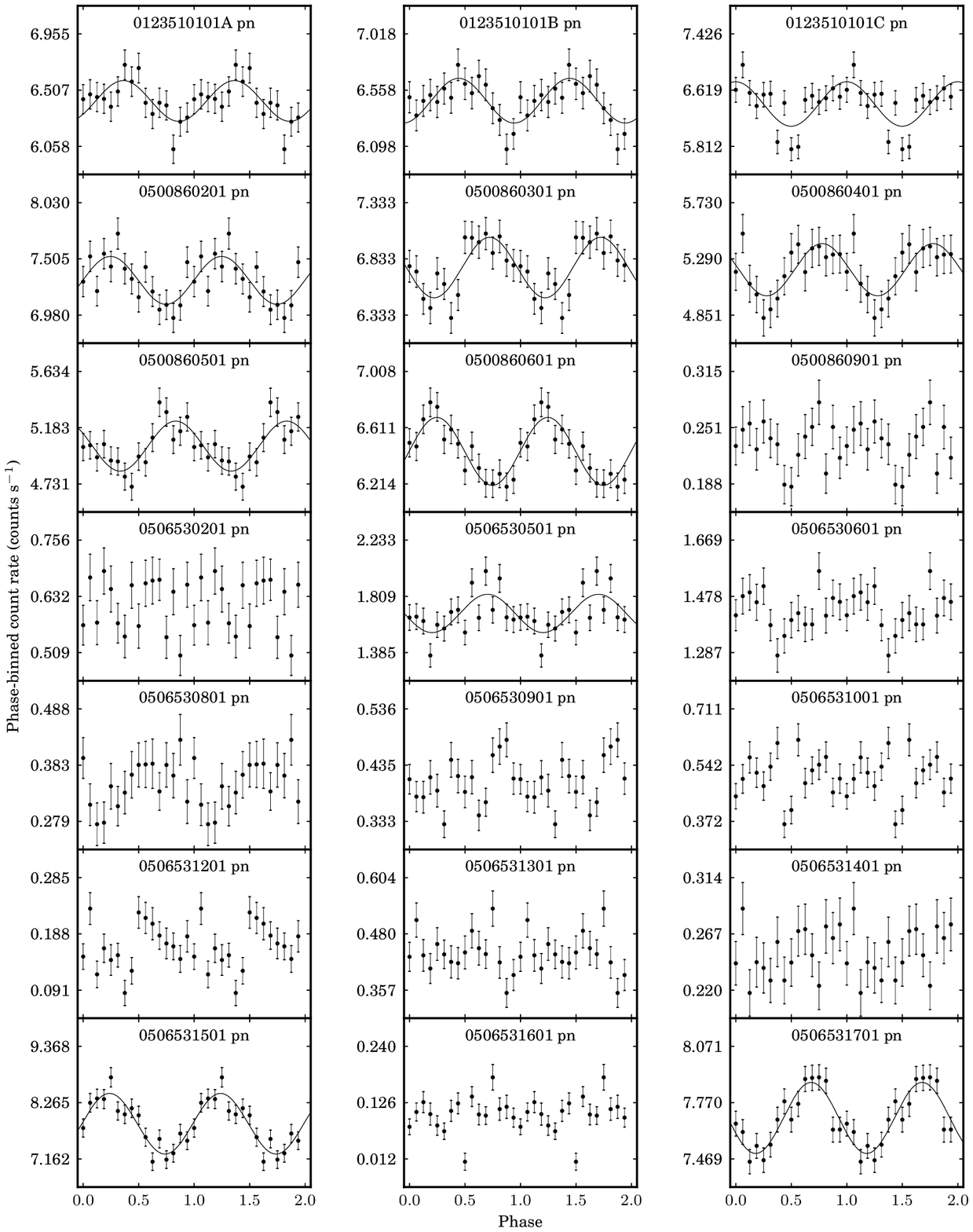}
  \caption{The EPIC pn broadband light curves of CAL~83 folded on the $\sim$67~s period. For the datasets in which the periodicity could be detected in the periodograms (see Fig.~\ref{fig:pnFIN005aP2H_scrgle}), the best fitting sine curve with the period fixed to the corresponding value provided in Table~\ref{tab:CAL83_XMM-Newton_all_obs} is overplotted. The other 10 datasets were folded on the mean period of 67~s. The starting point for the phase folding was taken at the start of each dataset. Two cycles are plotted for clarity.}
    \label{fig:pnFIN005aH_fold}
\end{figure*}

The 11 EPIC light curves exhibiting $P_{67}$ were subsequently folded on their corresponding periods as given in Table~\ref{tab:CAL83_XMM-Newton_all_obs}, using the starting point of each light curve as the reference point. For each of these, a sine curve was also fitted to the folded light curve, keeping the period fixed to the value in Table~\ref{tab:CAL83_XMM-Newton_all_obs}. The semi-amplitude of the modulation was calculated by expressing the semi-amplitude of the sine fit to the EPIC pn data as a percentage of the mean of the fit, and is also given in Table~\ref{tab:CAL83_XMM-Newton_all_obs}. The folded EPIC pn light curves are presented in Fig.~\ref{fig:pnFIN005aH_fold}. One can see that most of these do indeed follow a quasi-sinusoidal modulation pattern, with the semi-amplitude ranging from 2.5 to 8.6~per~cent. (The folded pn light curves of some of these observations were presented in figure 3 of \citealp{Odendaal_etal2014a}, together with the MOS light curves where $P_{67}$ was evident, but here the full set of pn light curves is provided, also incorporating the correction to thin filter counts.)

The other 10 datasets where the periodicity was not detected in the periodograms, were also folded, but on the mean period of 67~s, as an additional check of whether they exhibit any traces of such a modulation. No sine curves were fitted for these. In several of these datasets, a $\sim$67~s modulation may indeed be present in the data, even if not at a very significant level. One must also keep in mind that, if a similar modulation is inherently contained in these observations, it may well be at a period slightly different from the $\sim$67~s mean, which will change the appearance of the folded light curve. 

As an additional investigation of the possible occurrence of the same periodicity in the MOS data, each MOS light curve was folded on the same period as the corresponding pn observation, using the same reference point as the pn light curves so that their phasing can be compared. For the 11 datasets containing $P_{67}$, the folded pn and MOS light curves were found to be in phase, with the exception of 0123510101B, where the MOS light curve was very noisy. 

It is quite obvious that there is considerable variability in both the period and the amplitude of the periodicity in different observations. Within a single observation, the peak often seems to be multiperiodic (see Fig.~\ref{fig:pnFIN005aP2H_scrgle}, especially observation 0506531701). By using Eq.~(\ref{eq:DeltaP}), it was determined that, to obtain a $\pm$1~s error in a measurement of a period with a value near 67~s, the length of the light curve needs to be 2245~s. To investigate this variability within a single observation, the EPIC pn light curves containing the periodicity were then divided into a series of consecutive segments, each with a length of exactly 2245~s. The starting point of a particular segment was displaced by a value close to 600~s relative to the starting point of the previous segment, with the exact displacement determined by the total length of the observation, so that all the data points in the observation could be used. These overlapping segments enabled the calculation of a 
`moving average' of the period through the course of the observation. A LS analysis and accompanying significance computation for frequencies between 10 and 20~mHz were performed for each segment. 

\begin{figure*} 
 \centering
 \includegraphics{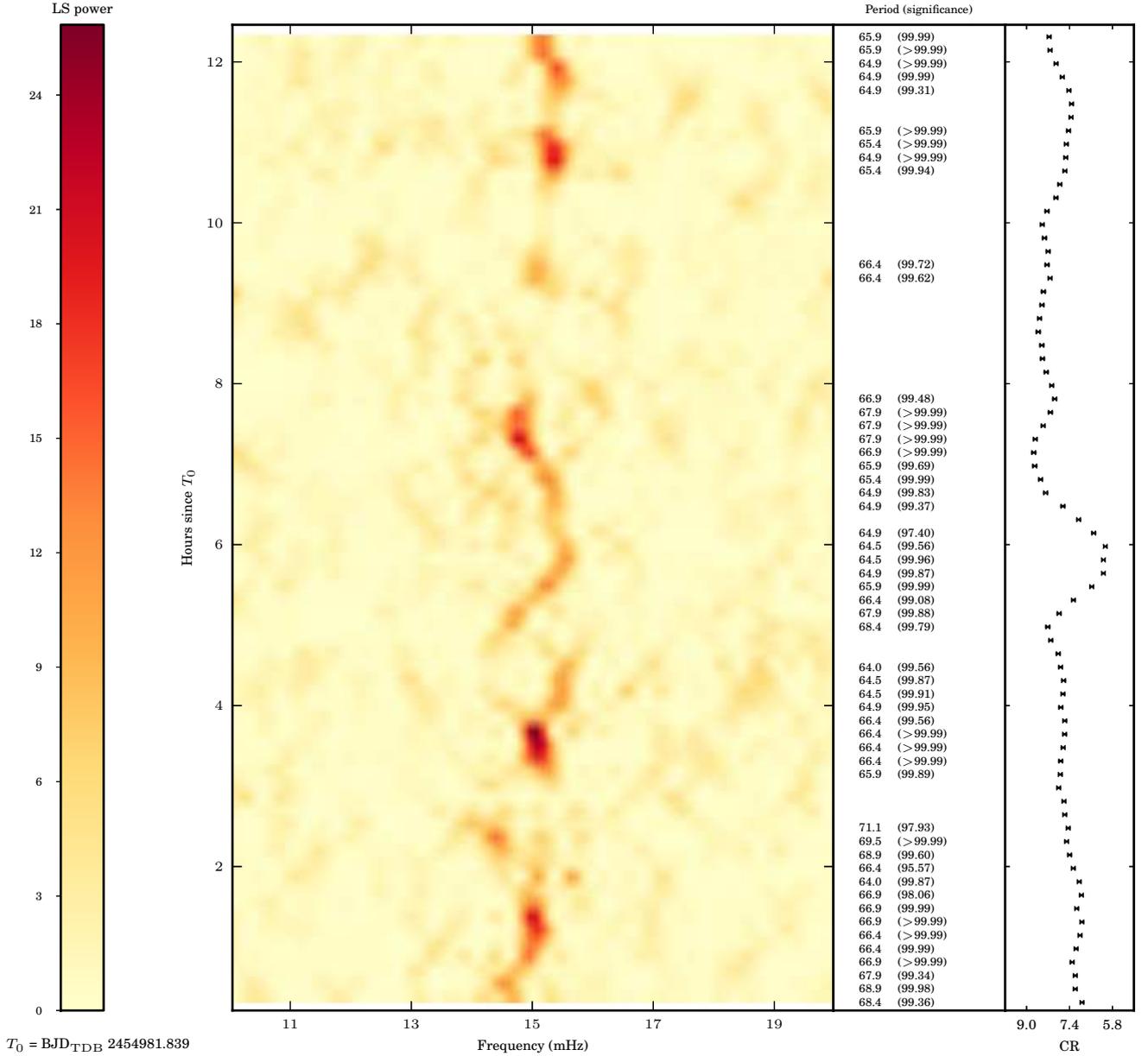}
 \caption{Lomb-Scargle (LS) periodograms of EPIC pn dataset 0506531701 of CAL~83 (the longest continuous \textit{XMM-Newton} observation), illustrating the variability in the value of the $\sim$67~s period. The colour represents the LS power, and in cases where a peak was detected between 13.5 and 16.5~mHz at a $>$95.45~per~cent significance level, the corresponding period in seconds and its significance percentage are given. The error in the period values is $\pm$1~s. The BJD$_{\rm TDB}$ reference represents the start of the observation. For a comparison with the X-ray count rate (CR), the broadband light curve is also provided.}
 \label{fig:dynamic_LS_0506531701}
\end{figure*}

The dynamical periodogram created in this way is shown in Fig.~\ref{fig:dynamic_LS_0506531701} for the EPIC pn light curve of the longest observation, 0506531701, containing 73 segments. The average CR per segment is also shown for comparison. The segments in which a period with frequency between 13.5~mHz and 16.5~mHz was found at a $>$95.45~per~cent level are annotated with the period and its significance. The nature of $P_{67}$ in the dynamical periodograms of the shorter observations was similar to that in 0506531701. 

It is evident that there is significant variability of at least $\sim$3~s to each side of the mean $P_{67}$, even on time-scales of a few hours, and that $P_{67}$ also tends to disappear and reappear again after some time. The length of observation 0506531701 is approximately one half of the orbital period, and if there was a direct correlation between the period value and the orbital motion, one would expect the period variation to take the approximate shape of one half of a sinusoid. However, the pattern of the variability of the modulation in this observation does not suggest any correlation whatsoever with the orbital motion. At first glance, Fig.~\ref{fig:dynamic_LS_0506531701} seems to suggest that there may be a correlation between the period value and the CR, but a correlation study did not reveal a statistically significant correlation.

As explained in \S\ref{sec:CV_quasi-periodic_modulations}, the quality factor $Q$ can be used to quantify the coherence of a periodicity. To quantify the variability in $P_{67}$, the period values and corresponding times of observation $t$ were extracted from observation 0506531701, utilizing data only from the segments in Fig.~\ref{fig:dynamic_LS_0506531701} with a $>$95.45~per~cent detection significance. The value of $Q$ was evaluated for every set of two measurements $(t_i,P_i)$ and $(t_{i+1}, P_{i+1})$ in the light curve (provided that $P_i\neq P_{i+1}$), using
\begin{equation}
 Q = \left| \frac{P_{i+1}-P_i}{t_{i+1} - t_i} \right|^{-1}~,
\end{equation}
yielding a value for $Q$ at time $(t_i + t_{i+1})/2$. Standard propagation of the errors in the periods were used to obtain the errors in $Q$. Considering only $Q$-values for which the error bars did not include zero, the resultant $Q$-values were between $210\pm100$ and $460\pm120$, with a mean of $343\pm87$. The second-longest continuous observation, 0500860601, yielded similar results.

\cite{Odendaal_etal2014a} also investigated the possibility that $P_{67}$ might represent the beat period between a very short WD rotation period and the Keplerian period of dense blobs in the inner accretion disc close to the WD. In such a scenario, the latter period might only be quasi-periodic, due to variations in the Keplerian radius and associated Keplerian period of such orbiting inhomogeneities, yielding a variable beat period. \cite{Odendaal_etal2014a} showed that a 67~s beat period could in principle result from a Keplerian period in the $\sim$4--15~s range, and a WD rotation period in the $\sim$4--12~s range (just above the break-up period of the WD), but no consistent X-ray modulation was found at such high frequencies during the timing analysis. The eLIMA model described in the next section provides a more plausible framework for the interpretation of $P_{67}$.

\section{The \lowercase{e}LIMA model for CAL~83}
\label{sec:eLIMA_CAL83}

The properties of the observed $P_{67}$ are very similar to those of DNOs. The observed modulation amplitudes fall in the range that has previously been observed for DNO light curves, although $P_{67}$ seems to be slightly less coherent (with $Q$ of the order of $10^2$) than typically expected for DNOs (with $Q$ usually in the $10^3$--$10^7$ range). This leads one to consider the possibility that, like DNOs, $P_{67}$ may originate in an equatorial belt around the WD. The lower coherence of $P_{67}$ can be expected, due to the more dynamic nature of the environment on the WD surface in SSSs. With the nature of DNOs in dwarf novae being explained by the well-established LIMA model (summarized in \S\ref{sec:LIMA_DNe}), we now propose the application of a similar model to the more `extreme' environment on the surface of the WD in SSS CAL~83, i.e.~the eLIMA model.

In this section, it is illustrated in a broad quantitative manner that the eLIMA model for $P_{67}$ is indeed plausible. First, some context is provided for this model in CAL~83 by some remarks on the belt location in \S\ref{subsec:belt_location}, followed in \S\ref{subsec:WD_magnetic_field} by a discussion of the magnitude of the WD magnetic field, and its implications regarding the flow of material in the vicinity of the belt. The mass of the equatorial belt is estimated in \S\ref{subsec:belt_mass}. The generation of a toroidal magnetic field within the rotating belt is discussed in \S\ref{subsec:toroidal_field}, as well as a potential explanation of the variations in $P_{67}$ within the eLIMA framework. The section concludes with the investigation of possible longer time-scale X-ray periodicities related to $P_{67}$ in \S\ref{subsec:longer_X-ray}.

\subsection{The location of the equatorial belt}
\label{subsec:belt_location}

According to \cite{Lanz_etal2005a}, the photospheric radius of the WD in CAL~83 is $R_{\rm phot}\sim(7.0\pm0.7)\times10^8{\rm~cm}$, and this will be regarded as an approximate outer limit for the equatorial belt. \citeauthor{Lanz_etal2005a} estimated the WD mass as $\sim$1.3${\rm~M}_\odot$, and the resulting Keplerian radius associated with a period of 67~s is
\begin{equation}
  R_{\rm K} = 2.7 \times 10^9 \left( \frac{M_1}{1.3{\rm~M}_\odot} \right)^{1/3} \left( \frac{P_{\rm K}}{67{\rm~s}} \right)^{2/3} {\rm~cm}~,
\end{equation}
while the Keplerian period associated with $R_{\rm phot}$ is significantly smaller, i.e.
\begin{equation}
  P_{\rm K} \approx 9 \left( \frac{M_1}{1.3{\rm~M}_\odot} \right)^{-1/2} \left( \frac{R_{\rm phot}}{7.0\times10^8{\rm~cm}} \right)^{3/2}{\rm~s}~.
\end{equation}
Therefore, an equatorial belt with rotation period $\sim$67~s at $\sim$7$\times10^8{\rm~cm}$ rotates more slowly than the local Keplerian velocity, possibly due to the magnetic coupling of the belt with a WD core that has an even longer rotation period.

\subsection{The magnetic field of the white dwarf}
\label{subsec:WD_magnetic_field}

If the belt now has a larger angular velocity than the WD core, it indicates that the WD magnetic field is too weak to rigidly couple the belt to the core. An upper limit to the magnetic field strength can be calculated with Eq.~(\ref{eq:Katz_B}), but an estimate of the potential spin-up rate $\dot{\Omega}_*$ as a result of accretion torques must first be obtained. To perform this estimate, one needs to consider the mass, radius and magnetic field of the WD.

One must keep in mind that $R_{\rm phot}$ includes not only the WD core, but also an extended accreted envelope on the WD surface, with hydrogen burning at its base. In order to determine the actual radius of the WD core, one can consider the radius $R_1$ of a zero-temperature WD that is related to its mass $M_1$ by (\citealp{hamada_salpeter1961,eracleous_horne1996})
\begin{equation}
  \label{eq:hamada_salpeter}
  R_1=4.0\times10^8\left(\frac{M_1}{1.3{\rm~M}_\odot}\right)^{-0.8}{\rm~cm}~.
\end{equation}
The WD radius derived with Eq.~(\ref{eq:hamada_salpeter}) can be considered as a lower limit when considering hot WDs, as shown by \cite*{Panei_etal2000}. Therefore the radius of the WD in CAL~83 can be estimated as $R_1\gtrsim4.0\times10^8{\rm~cm}$. At a certain distance from the WD, the magnetic pressure will be equal to the ram and gas pressure of the accreting material. This limiting radius is known as the Alfv\'en radius. For a spherical accretion flow, the Alfv\'en radius is given by \citep{ElsnerLamb1977}
\begin{equation}
  r_{\rm M} = \left( \frac{B_1^4 R_1^{12}}{8 G M_1 \dot{m}_{\rm acc}^2} \right)^{1/7}~,
  \label{eq:spherical_Alfven}
\end{equation}
where $B_1$ represents the dipole field strength of the WD. For a cylindrical configuration, the Alfv\'en radius (also known as the magnetospheric radius) is typically given by \citep{GhoshLamb1979b}
\begin{equation}
 R_{\rm M} \sim 0.5 r_{\rm M}~.
 \label{eq:cyl_Alfven_small}
\end{equation}
If $R_{\rm M}>R_1$, the accretion disc structure will be disrupted, and for radii $R<R_{\rm M}$, the accreting material will flow along the magnetic field lines, and be channelled onto the polar caps of the WD. For example, in the case of CAL~83, if we assume $ \dot{m}_{\rm acc}\sim10^{-7}{\rm~M}_\odot{\rm~yr}^{-1}$, and if $B_1\sim10^5{\rm~G}$ (see below), Eq.~(\ref{eq:spherical_Alfven}) yields
\begin{align}
 r_{\rm M} &\sim 2 \times 10^8 \left( \frac{B_1}{10^5{\rm~G}} \right)^{4/7} \left( \frac{R_1}{4\times10^8{\rm~cm}} \right)^{12/7} \times\nonumber\\
 &~~~~~\left( \frac{M_1}{1.3{\rm~M}_\odot} \right)^{-1/7} \left( \frac{\dot{m}_{\rm acc}}{10^{-7}{\rm~M}_\odot{\rm~yr}^{-1}} \right)^{-2/7}{\rm~cm}~,
 \label{eq:r_M_CAL83}
\end{align}
implying $R_{\rm M}\sim10^8{\rm~cm}$ according to Eq.~(\ref{eq:cyl_Alfven_small}). Considering a case where the WD rotates as a rigid body, and experiences `spin-up' due to the transfer of angular momentum from the accretion disc, the rate of change of the spin angular velocity of the WD can be expressed as (e.g.~\citealp[pp.~162--163]{frank_etal2002}):
\begin{equation}
\label{eq:dot_Omega_*}
 \dot{\Omega}_* = \dot{m}_{\rm acc} \left( G M_1 R_{\rm M}\right)^{1/2} I^{-1}~,
\end{equation}
where $I$ is the moment of inertia of the WD. For a solid spherical body, $I=\frac{2}{5}M_1 R_1^2$, i.e.~in CAL~83,
\begin{equation}
 I \sim 1.6\times10^{50} \left( \frac{M_1}{1.3{\rm~M}_\odot} \right) \left( \frac{R_1}{4.0\times10^8{\rm~cm}} \right)^2 {\rm~g~cm}^2~.
\end{equation}
One can see that the value of $\dot{\Omega}_*$ is proportional to $B_1$ to the power $2/7$ (through its dependence on $R_{\rm M}$), and therefore does not change too drastically with large changes in magnetic field. E.g.,  while keeping the other parameters the same as previously, the value of $\dot{\Omega}_*$ changes by only one order of magnitude from $10^{-15}{\rm~s}^{-2}$ to $10^{-14}{\rm~s}^{-2}$ when varying the dipole field strength from $B_1 \sim 3\times 10^2{\rm~G}$ to $\sim$10$^6{\rm~G}$.

Once again assuming $\rho_{\rm wd}\sim10^6{\rm~g~cm}^{-3}$, the upper limit for the product of the interior magnetic field components to avoid rigid rotation can now be calculated with Eq.~(\ref{eq:Katz_B}). When varying $\dot{\Omega}_*$ between $10^{-15}{\rm~s}^{-2}$ and $10^{-14}{\rm~s}^{-2}$, and the radius between $R_1\sim4\times10^8{\rm~cm}$ and $R_{\rm phot}\sim7\times10^8{\rm~cm}$, the value of $B_r\sim B_\phi$ varies from a few times $10^4{\rm~G}$ to a few times $10^5{\rm~G}$. In turn, the inner field strengths can be considered as an upper limit to the strength of the surface field $B_1$. As an order of magnitude estimate, the primary field strength should therefore obey the limit $B_1\lesssim10^5{\rm~G}$ to allow the existence of a non-corotating belt structure at the inner edge of the disc.

\subsection{The mass of the equatorial belt}
\label{subsec:belt_mass}

If the observed decreases in $P_{67}$ result from spin-up of the equatorial belt, this places certain constraints on the moment of inertia and thus the mass of the belt. The maximum spin-up rate $\dot{\Omega}_{\rm b}$ of the belt can be estimated by considering the period variations as illustrated in Fig.~\ref{fig:dynamic_LS_0506531701}. The rapid decrease in $P_{67}$ from 5 to 5.8 hours elapsed since $T_0$ in observation 0506531701 will be assumed to be representative of the most rapid change in the period of the equatorial belt ($P_{\rm b}$), and will be used to estimate the maximum value of $\dot{\Omega}_{\rm b}$. The period difference between the two segments is $\Delta P = (68.4 - 64.5){\rm~s} =3.9{\rm~s}$, and the time elapsed between them is $\Delta t = 3000{\rm~s}$. Thus, $\dot{P}_{\rm b} = \Delta P / \Delta t = 1.3\times10^{-3}$, and since $\Omega_{\rm b}=2\pi /P_{\rm b}$,
\begin{equation}
 \dot{\Omega}_{\rm b} \sim 1.8 \times 10^{-6} \left( \frac{\dot{P}_{\rm b}}{1.3\times10^{-3}} \right) \left( \frac{P_{\rm b}}{67{\rm~s}} \right)^{-2}  {\rm~s}^{-2}~.
\end{equation}
In Eq.~(\ref{eq:dot_Omega_*}), the factor $( G M_1 R_{\rm M})^{1/2}$ represents specific angular momentum in a Keplerian rotating mass element. The specific angular momentum of the belt magnetically linked to the WD is estimated by $Rv_{{\rm b},\phi}$, where $v_{{\rm b},\phi}=2\pi R/P_{\rm b}$ is its azimuthal velocity. The spin-up rate $\dot{\Omega}_{\rm b}$ of the belt can thus be expressed as 
\begin{align}
 \dot{\Omega}_{\rm b} = \dot{m}_{\rm acc}\left( \frac{2\pi R^2}{P_{\rm b}} \right)~I_{\rm b}^{-1}~,
\end{align}
or, solving for the moment of inertia $I_{\rm b}$ of the belt, one obtains
\begin{align}
  I_{\rm b} &= \frac{\dot{m}_{\rm acc}}{\dot{\Omega}_{\rm b}}\left( \frac{2\pi R^2}{P_{\rm b}} \right) \nonumber \\
	    &\sim 1.6\times10^{41} \left( \frac{\dot{m}_{\rm acc}}{10^{-7}{\rm~M}_\odot{\rm~yr}^{-1}} \right) \left( \frac{\dot{\Omega}_{\rm b}}{1.8 \times 10^{-6}{\rm~s}^{-2}} \right)^{-1} \times\nonumber\\
	    &~~~\left( \frac{R}{7.0\times10^8{\rm~cm}} \right)^{2} \left( \frac{P_{\rm b}}{67{\rm~s}} \right)^{-1}   {\rm~g~cm}^2~,
\end{align}
which is $\sim$9 orders of magnitude smaller than that of the WD. This represents a maximum value for the inertia \textit{if} the observed $\dot{\Omega}_{\rm b}$ is to be explained by the spin-up of the belt. One can now proceed to estimate the mass of the belt. For simplicity, it will be assumed that the geometry of the belt can be approximated as an annular cylinder around the WD, with central axis perpendicular to the orbital plane. The outer radius $R_{\rm b2}$ of the cylinder will be taken as approximately equal to $R_{\rm phot}\sim7\times10^8{\rm~cm}$. The inner radius $R_{\rm b1}$ of the cylinder can be set to the minimum WD radius ($\sim$4$\times10^8{\rm~cm}$), although it is expected to be larger than this. The moment of inertia is then given by (e.g.~\citealp*[p.~253]{Halliday_etal2005})
\begin{equation}
 I_{\rm b} = \frac{1}{2}m_{\rm b}\left(R_{\rm b1}^2 + R_{\rm b2}^2\right)~,
\end{equation}
where $m_{\rm b}$ is the mass of the belt, which can be solved for as
\begin{equation}
 m_{\rm b} = \frac{2I_{\rm b}}{R_{\rm b1}^2 + R_{\rm b2}^2}= 2.5\times10^{-10}{\rm~M}_\odot
\end{equation}
This can now be compared to the expected mass of the accreted envelope on the WD surface. Nuclear fusion can be ignited in an envelope of accreted hydrogen-rich material on the surface of the WD if a critical envelope mass $m_{\rm env}^{\rm crit}$ is reached, in which case the temperature and pressure conditions required for the nuclear burning of hydrogen to helium by the CNO cycle can be sustained, i.e.~$T\sim 10^8{\rm~K}$ and $P\gtrsim 10^{18}$--$10^{20}\text{~dyn}{\rm~cm}^{-2}$ \citep{Fujimoto1982}. The mass $m_{\rm env}^{\rm crit}$ is given by
\begin{align}
\label{eq:critical_env_mass}
 \log \left( \frac{m_{\rm env}^{\rm crit}}{{\rm M}_\odot} \right) \approx&-2.862 + 1.542 \left( \frac{M_1}{{\rm M}_\odot} \right)^{-1.436} \times\nonumber\\
 &\ln \left( 1.429 - \frac{M_1}{{\rm M}_\odot} \right) - \nonumber\\
 &0.197 \left[ \log\left(\frac{\dot{m}_{\rm acc}}{{\rm M}_\odot{\rm~yr}^{-1}}\right) +10\right]^{1.484}
\end{align}
(see also \citealp{PrialnikKovetz1995,TownsleyBildsten2004}). The higher the WD mass and/or the accretion rate, the lower the required value of $m_{\rm env}^{\rm crit}$. For a $1.3{\rm~M}_\odot$ WD accreting at $10^{-7}{\rm~M}_\odot{\rm~yr}^{-1}$, one obtains $m_{\rm env}^{\rm crit}\sim9\times10^{-7}{\rm~M}_\odot$, which can be considered as a lower limit for the envelope mass for nuclear burning to occur. Therefore, if $m_{\rm b}=2.5\times10^{-10}{\rm~M}_\odot$, the equatorial belt represents only a small fraction of the total envelope mass.

Considering the extended, `fuzzy' nature of the WD envelope in SSSs, cloaking the nuclear burning shell, this scenario seems quite plausible. An equatorial belt at the envelope-disc boundary may have a rotational period of $\sim$67~s as described above, while accreted layers closer to the WD may be rotating with slightly longer periods, yielding differential rotation between these weakly coupled layers at different radial distances. In such a case it is more likely that the inner radius of the belt is significantly larger than the WD radius. Increasing $R_{\rm b1}$ up to almost $R_{\rm b2}=7\times10^8{\rm~cm}$ (implying a thin shell at this radius) yields $m_{\rm b}=1.6\times10^{-10}{\rm~M}_\odot$, i.e.~the presumed thickness of the shell does not change the resulting mass significantly.

\subsection{The toroidal field, and spin-up/spin-down of the belt}
\label{subsec:toroidal_field}

In \S\ref{subsec:WD_magnetic_field}, the magnetospheric radius $R_{\rm M}$ was estimated to be $\sim$10$^8{\rm~cm}$ for CAL~83 when $B_1$ is of the order of $10^5{\rm~G}$, i.e.~smaller than that of the WD and its extended envelope. If one would increase the value of $R_1$ to $R_{\rm phot}\sim7\times10^8{\rm~cm}$ in Eq.~(\ref{eq:spherical_Alfven}), one would find $R_{\rm M}\sim3\times10^8{\rm~cm}$, which is still smaller than the WD radius. However, from the dependence of $r_{\rm M}$ on $\dot{m}_{\rm acc}$ in Eq.~(\ref{eq:spherical_Alfven}), it is obvious that, for a WD with a given $B_1$, $R_1$ and $M_1$,  the much lower accretion rates in DNe (even during outburst) are always associated with larger magnetospheric radii than what would be found in SSSs. This subtle difference indicates that, for similar dipole field strengths, the high $\dot{m}_{\rm acc}$ in SSSs will cause the ram pressure to have a more pronounced influence close to the WD surface, compared to the magnetic pressure, than in dwarf novae. For DNe, this results in the usual situation of magnetically controlled accretion from the inner disc onto the equatorial belt, according to the LIMA model.

In CAL~83, the WD magneto\-sphere will be compressed much more strongly, so that the disc is expected to extend all the way down to the WD envelope, so that the equatorial belt in the eLIMA model can be considered as a hot, torus-like `boundary layer' where the inner disc blends into the WD envelope. One therefore has more or less Keplerian flow in the inner disc itself, with period $P_{\rm K}$, a much longer WD rotation period $P_*$, and a belt period such that $P_{\rm K} < P_b < P_*$. The belt rotation is slower than the local Keplerian velocity, due to its inertia, and especially its (weak) magnetic coupling to the central WD.

It is especially important to realize that the ram pressure dominated flow in the eLIMA model will result in a significant winding up of the WD magnetic field frozen into the belt plasma, resulting in the generation of a significant toroidal external field. Assuming that the rotation period of the WD is substantially longer than that of the belt, the belt will carry the foot-points of the field lines around its orbit, stretching the field lines and creating a substantial toroidal field component $B_{{\rm b},\phi}$ in the vicinity of the belt (e.g.~\citealp{AlyKuijpers1990}). It was shown by \cite{WarnerWoudt2002} that the toroidal field generated in the belt is given by
\begin{align}
 B_{{\rm b},\phi} \sim~&2\times10^5  \left( \frac{B_r}{10^5{\rm~G}} \right)^{1/2} \left( \frac{P_{\rm b}}{67{\rm~s}} \right)^{-1/2} \times \nonumber\\
 &\left( \frac{m_{\rm b}}{2\times10^{-10}{\rm~M}_\odot} \right)^{1/4} \left( \frac{T}{5.5\times10^5{\rm~K}} \right)^{1/4} \times \nonumber\\
 & \left( \frac{x}{0.1} \right)^{-1/4} \left( \frac{y}{0.01} \right)^{-1/4} \left( \frac{M_1}{1.3{\rm~M}_\odot} \right)^{-1/4} {\rm~G}~.
\end{align}
Following \cite{WarnerWoudt2002}, the height of the belt (in the direction perpendicular to the orbital plane) is given by $xR_1$. The quantity $yg$ represents the effective gravity in the belt, where $g$ is the surface gravity of the WD. For typical parameters of DNe, these authors found $x\sim0.1$ and $y\sim0.01$. Adopting  $T=550000{\rm~K}$ for CAL~83 according to the fits of \cite{Lanz_etal2005a}, as well as $m_{\rm b}\sim2\times10^{-10}{\rm~M}_\odot$ (an average value estimated by considering the calculations in \S\ref{subsec:belt_mass}), yields a toroidal field of $\sim$2$\times10^5{\rm~G}$ for an original poloidal field of $10^{5}{\rm~G}$, which indicates a significant enhancement.

It has also been shown (\citealp{GhoshLamb1991}, e.g.~\citealp{MeintjesDeJager2000}) that when the stretched toroidal field exceeds the poloidal WD field by up to 10 times, the field will become unstable to magnetic reconnection. The following mechanism is now proposed for the observed variability in $P_{67}$:

It is anticipated that a spin-up/spin-down behaviour of the belt modulates $P_{67}$.  We propose a cyclic variation of the magnetic coupling between the WD core and the belt.  It is anticipated that in the case of a very weak coupling, the belt is mainly spun up by accretion disc torques on the outside of the belt, resulting also in the gradual build-up of a significant toroidal field component.  This enhanced toroidal field will forge a stronger link between the slower rotating WD core and the belt, resulting in a subsequent nett spin-down effect.  For magnetic pitch angles $\gamma = B_{\rm b,\phi} / B_{{\rm b},r}>10$, the fields will become unstable to reconnection, with subsequent magnetic annihilation in unstable flux tubes.  This will weaken the magnetic connection with the slower rotating WD, resulting in a subsequent nett spin-up effect due to viscous effects of the accretion disc.

The belt is therefore simultaneously subject to both the spin-up effect of the inner disc and the spin-down effect of its `loose' magnetic coupling to the central WD, with the cyclic variations in the latter (due to toroidal field enhancement and reconnection) continually modulating the equilibrium value of the belt period, yielding the observed `wobble' in $P_{67}$.

In dwarf novae, the increase in the DNO period at the end of the outburst has been explained as a rapid spin-down of the equatorial belt following a decrease in the accretion rate. Although SSSs are not subject to the viscosity-related hysteresis curve of dwarf nova outbursts, they do undergo other processes that can modulate the mass transfer rate. However, these are unlikely to occur on such short time-scales, and the spin-down of the belt is much more likely to be due to the magnetic coupling as described above.

Alternatively, the rapid changes in the value of $P_{67}$ within a single observation may be caused by the same mechanism proposed to explain the discontinuous jumps in DNOs:  the channelling of the accretion flow onto different accretion arcs in the belt (at different latitudes with different associated rotation periods) as a result of magnetic reconnection. In such a case, the disappearance and reappearance of the period on time-scales of hours may be due to the accretion being channelled dominantly onto latitudes with the same period at some epochs (i.e.~a well-defined period is observable), with accretion onto a wide range of latitudes with different periods at other epochs (i.e.~a well-defined oscillation peak is not observable). Although the magnetospheric radius for a dipole field configuration is much smaller in SSSs than in DNe, the significant enhancement of the toroidal field may cause the magnetic loops associated with such accretion arcs to be `visible' in the optically thin outer regions of the fuzzy WD envelope, and have an enhanced influence on the accretion trajectories of the accreting material. 

If spin-up/spin-down is not the primary cause of the period modulations, then the upper limit imposed on the inertia earlier in this section would not be so strict, and a larger fraction of the envelope mass could form part of the equatorial belt. However, it is probable that the variation in $P_{67}$ might be caused by a combination of spin-up/spin-down and accretion channel modulations.

Comparing the mean period and X-ray luminosity from observation to observation, marginal evidence has been found that $P_{67}$ is on average somewhat shorter when the X-ray luminosity (and the corresponding temperature) is higher. This could simply be due to the fact that higher CRs and temperatures indicate a slightly more contracted photospheric radius, with the belt at the envelope-disc boundary subsequently being at a slightly smaller radius with a smaller corresponding period. However, it is noted that the modulation in $P_{67}$ from observation to observation over the years is not any larger than its modulation on time-scales of hours within a single observation.

\subsection{Longer X-ray periodicities possibly related to 67~s?}
\label{subsec:longer_X-ray}

\begin{figure*}
  \includegraphics{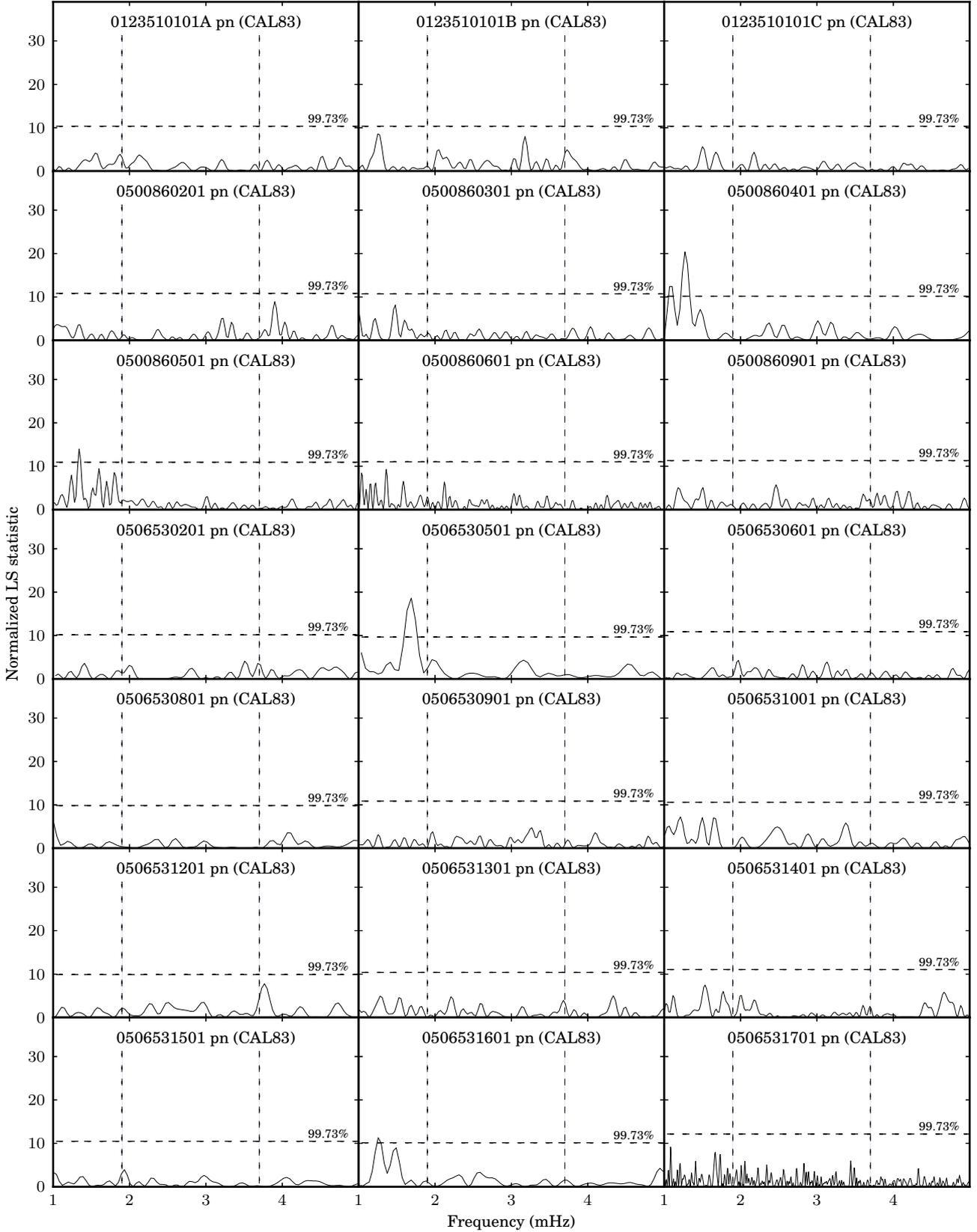}
\caption{Lomb-Scargle (LS) periodograms in the 1--5~mHz range of the EPIC pn data of CAL~83. The dashed vertical lines indicate the position of 1.9 and 3.7~mHz respectively.}
\label{fig:pnFIN010aP2M_scrgle}
\end{figure*}

\begin{figure}
\centering
  \includegraphics{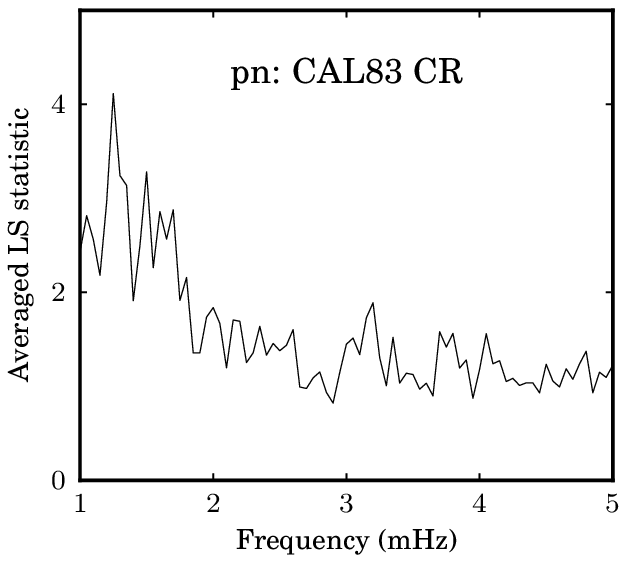}
\caption{The averaged Lomb-Scargle (LS) periodogram based on the EPIC pn broadband count rate (CR), in the 1--5~mHz range. No prominent peaks at 1.9 and 3.7~mHz are apparent.}
\label{fig:avgLSper_pnFIN005aP2HI1_5_mHz}
\end{figure}

As discussed in \S\ref{sec:LIMA_DNe}, DNOs in DNe are often observed concurrently with lpDNOs, which are also related to the WD spin period $P_*$ in terms of $P_* \sim 2 P_{\rm lpDNO} \sim 8 P_{\rm DNO}$. If $P_{67}$ in CAL~83 is analogous to an `ordinary' DNO in a dwarf nova, one could hypothetically expect the inherent presence of additional oscillations close to $P_{\rm lpDNO}=4\times67{\rm~s}=268{\rm~s}$ (3.7~mHz) and $P_*=8\times67{\rm~s}=536{\rm~s}$ (1.9~mHz) in the light curves (although the presence of a DNO does not necessarily mean that the lpDNO and/or WD rotation period should also be detectable). The results of a LS analysis in the 1--5~mHz region of the CAL~83 EPIC pn light curves with 10~s binning and detrended by subtracting a 2$^\text{nd}$ order polynomial fit, are shown in Fig.~\ref{fig:pnFIN010aP2M_scrgle}. An average periodogram of CAL~83 in the 1--5~mHz range, obtained by averaging all the EPIC pn on-state observations after using equal frequency bins during the LS analysis, is also shown in Fig.~\ref{fig:avgLSper_pnFIN005aP2HI1_5_mHz}.

The only peak close to either 3.7 or 1.9~mHz that has a $>$99.73\% significance level, is the one at $\sim$1.7~mHz (580~s) in observation 0506530501. However, this $\sim$1.7~mHz peak did not appear significant when using larger light curve bin sizes. However, it may be worthy of further investigation. None of the other observations contained an apparent peak at $\sim$1.9~mHz. In observation 0123510101B and especially 0500860201 and 0506531201, a weak peak is visible approximately at 3.7~mHz, although below the 99.73\% significance level. Although we can thus not confirm or exclude the presence of consistent periodicities at $P_{\rm lpDNO}=4\times67{\rm~s}$ and $2\times P_{\rm lpDNO}$, this is perhaps something that could be investigated in more detail using new X-ray data.

\section{SHOC optical light curves}
\label{sec:SHOC}

Optical light curves of CAL~83 were obtained with the Sutherland High-speed Optical Camera (SHOC) on the SAAO 1.9-m Telescope \citep{Coppejans_etal2013}, using a clear filter, on 8 nights during April 2013 and December 2014 (see Table~\ref{tab:SHOC_observations} for details). A comparison to the OGLE-IV\footnote{Fourth phase of the Optical Gravitational Lensing Experiment} light curve of CAL~83 showed that all these observations were performed while the source was in an optical low state (see \citealp[Chapter 6]{Odendaal2015} for more details).

The data reductions were performed according to standard {\sc iraf} procedures, as implemented in the Python- and PyRAF-based SHOC data reduction pipeline developed by M.~Kotze at the SAAO (the version updated on 3 June 2014), with minor adaptations by A.~Odendaal. Subsequent differential corrections using 4 comparison stars were applied, utilizing a method largely based on the one described by \cite{EverettHowell2001}. The corrected light curves are shown in Fig.~\ref{fig:SHOC_CAL83_difflcs}.

%%%%%%%%%%%%%%%%%%%%%%%%%%%%%%%%%%%%%%%%%%%%%%%%%%

\begin{table*}
\caption{SHOC observation sequences of CAL~83, together with instrumental configurations. Conventional (CON) mode was utilized for all observations, using a clear filter in both filter wheels.}
\label{tab:SHOC_observations}
  \begin{tabular}{@{}clrl@{~}c@{~}r@{~}c@{~}rccccc@{}}
  \hline
  Observation	&\multicolumn{2}{c}{Start date}		&\multicolumn{5}{c}{Exposures}	&Camera	&Pixel shift	&Gain	&Binning	&Observer\\
  number	&\multicolumn{2}{c}{\& time (UT)}	&&&&&				&	&speed (mHz)	&	&		&\\
  \hline
  01	&2013~Apr~10 	& 19:17	&1~s	&$\times$	&549	&=	&549~s	&SHOC2	&3 	&$5.2\times$	&$16\times16$		&A.~Odendaal\\
  02	&2013~Apr~11	& 20:01	&5~s	&$\times$	&718	&=	&3590~s	&SHOC2	&'' 	&''		&$8\times8$		&''\\
  03	&2013~Apr~12	& 17:50	&2~s	&$\times$	&2998	&=	&5996~s	&SHOC2	&'' 	&''		&''				&''\\
  04	&		& 19:32	&2~s	&$\times$	&2260	&=	&4520~s	&SHOC2	&'' 	&''		&''				&''\\
  05	&2013~Apr~13	& 17:42	&1.5~s	&$\times$	&2598	&=	&3897~s	&SHOC2	&'' 	&''		&''				&''\\
  06	&2013~Apr~14	& 18:28	&0.5~s	&$\times$	&3998	&=	&1999~s	&SHOC2	&'' 	&''		&''				&''\\
  07	&2013~Apr~15	& 17:53	&2~s	&$\times$	&1618	&=	&3236~s	&SHOC2	&'' 	&''		&''				&''\\
  08	&2014~Dec~16	& 20:43	&5~s	&$\times$	&298	&=	&1490~s	&SHOC1	&1	&$2.5\times$	&''				&L.~Klindt\\
  09	&		& 21:14	&10~s	&$\times$	&718	&=	&7180~s	&SHOC1	&''	&''		&''				&''\\
  10	&		& 23:20	&8~s	&$\times$	&718	&=	&5744~s	&SHOC1	&''	&''		&''				&''\\
  11	&2014~Dec~17	& 00:59	&8~s	&$\times$	&673	&=	&5384~s	&SHOC1	&''	&''		&''				&''\\
  \hline
  \end{tabular}
\end{table*}

%%%%%%%%%%%%%%%%%%%%%%%%%%%%%%%%%%%%%%%%%%%%%%%%%%

\begin{figure*}
 \centering
 \includegraphics{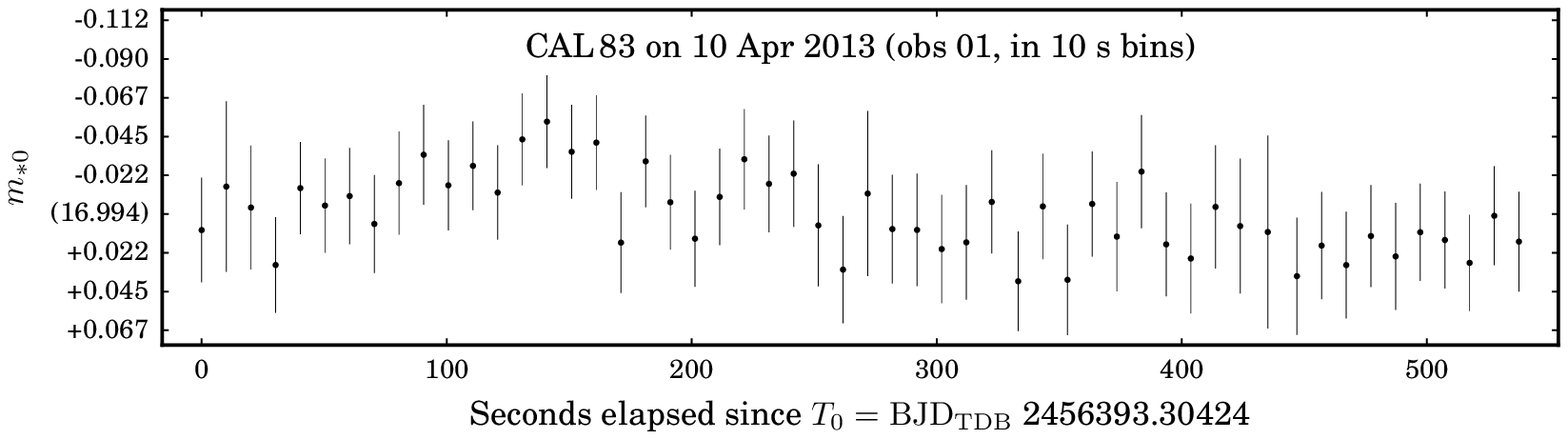}\\[15pt]
 \includegraphics{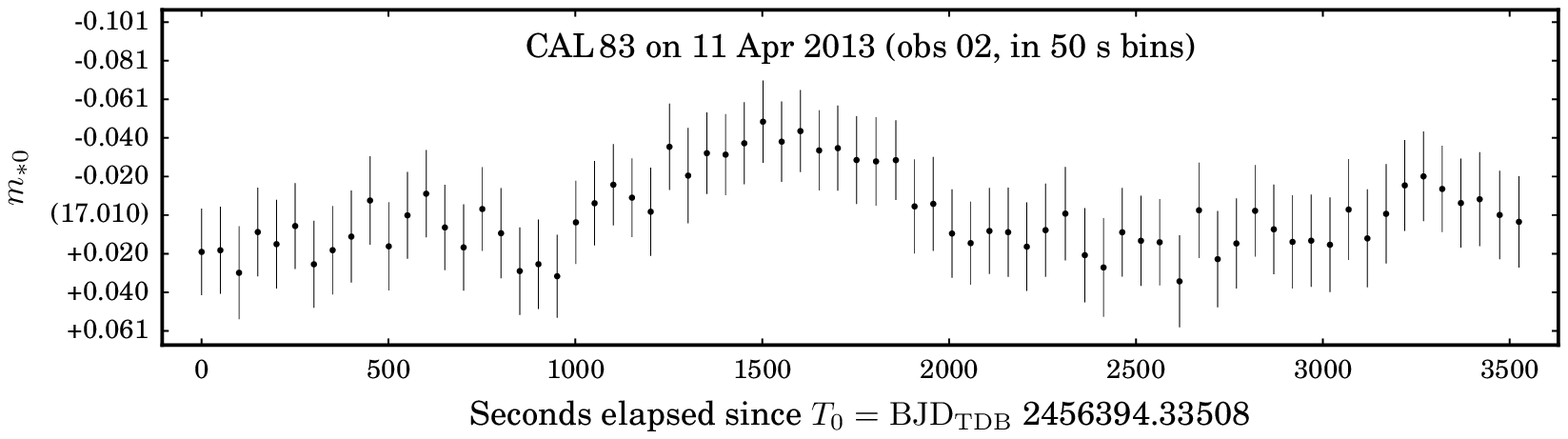}\\[15pt]
 \includegraphics{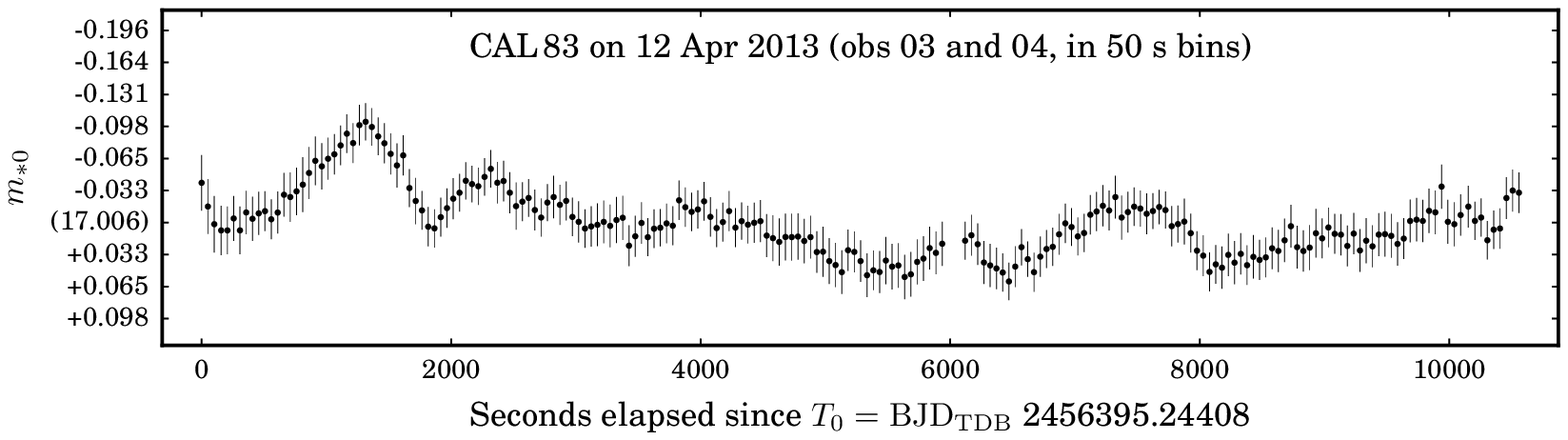}\\[15pt]
 \includegraphics{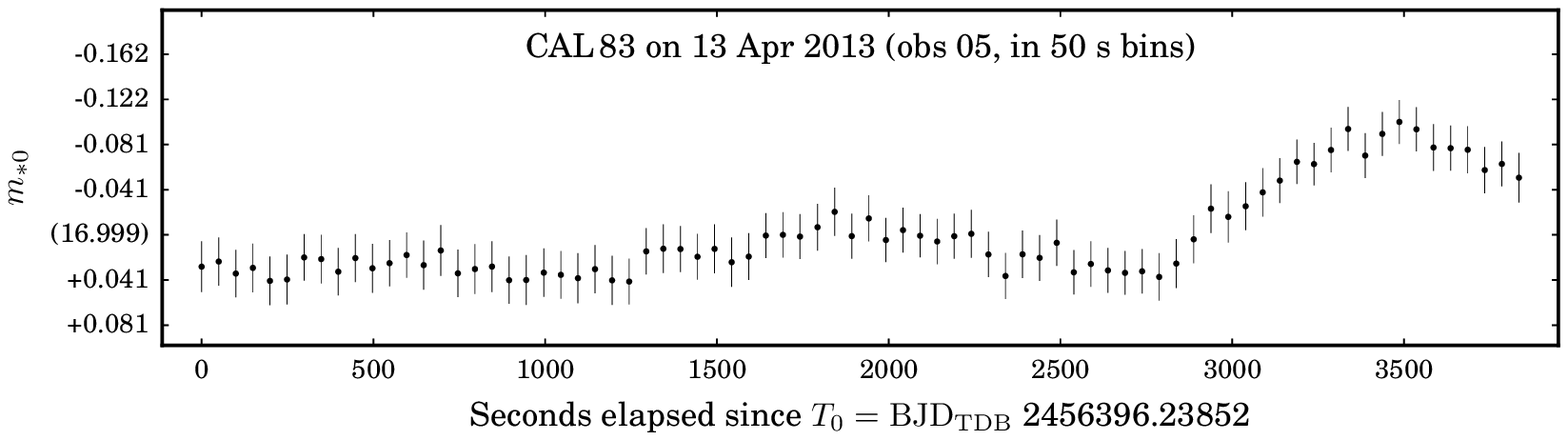}
 \caption{The SHOC light curves of CAL~83 after the differential correction. The horizontal axis indicates the seconds elapsed since the first measurement in the observation. On the vertical axis, the mean corrected instrumental magnitude is given in parentheses, with the other markings indicating integer multiples of the standard deviation of the light curve, referred to the mean value. The observation number in the respective labels refers to the number allocated in Table~\ref{tab:SHOC_observations}.}
 \label{fig:SHOC_CAL83_difflcs}
\end{figure*}

\addtocounter{figure}{-1}

\begin{figure*}
 \centering
 \includegraphics{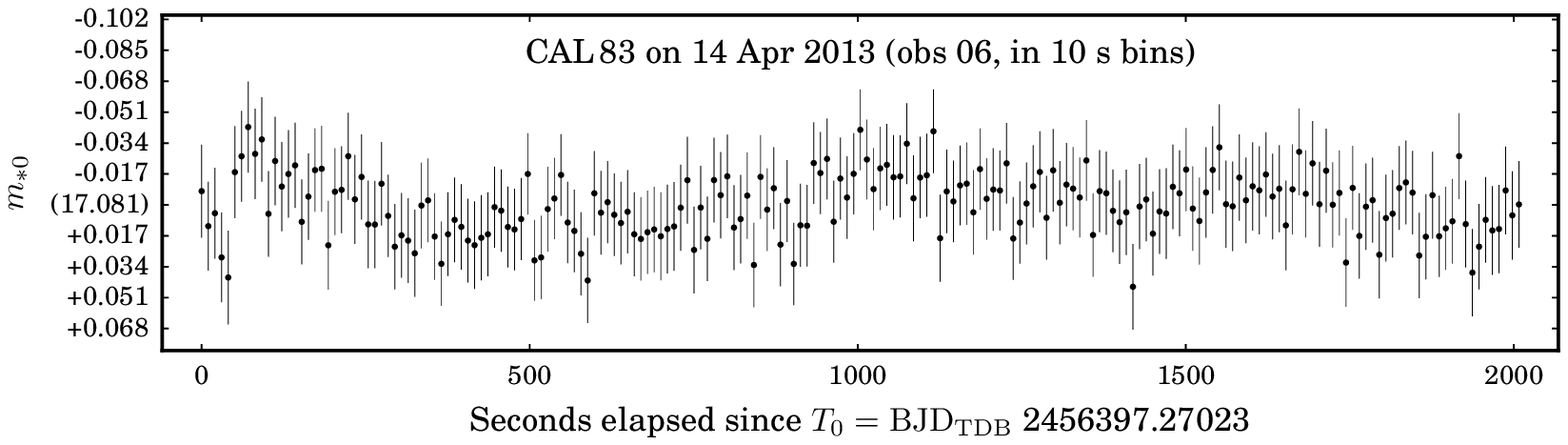}\\[15pt]
 \includegraphics{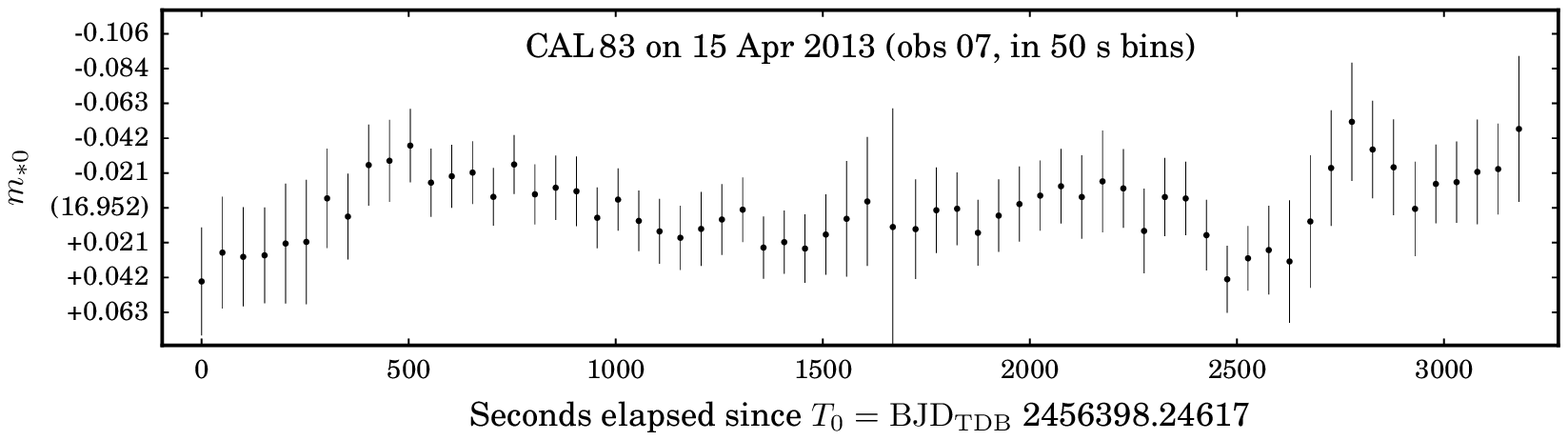}\\[15pt]
 \includegraphics{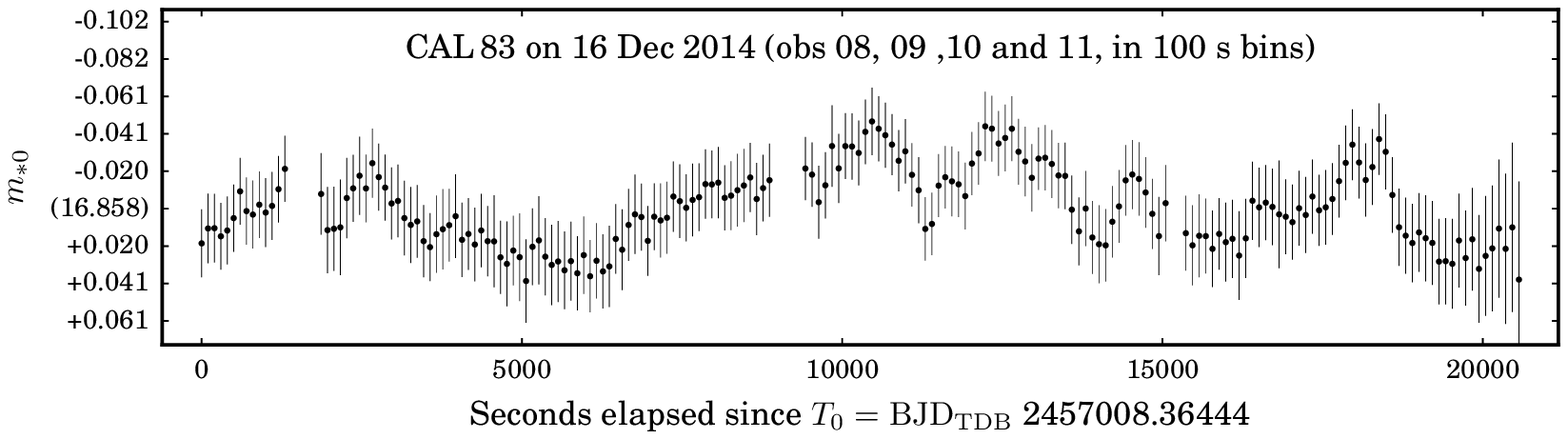}
 \caption{(continued) The SHOC light curves of CAL~83 after the differential correction. The horizontal axis indicates the seconds elapsed since the first measurement in the observation. On the vertical axis, the mean corrected instrumental magnitude is given in parentheses, with the other markings indicating integer multiples of the standard deviation of the light curve, referred to the mean value. The observation number in the respective labels refers to the number allocated in Table~\ref{tab:SHOC_observations}.}
\end{figure*}

%%%%%%%%%%%%%%%%%%%%%%%%%%%%%%%%%%%%%%%%%%%%%%%%%%

A LS analysis of the SHOC light curves did not reveal any discernible power peak around $\sim$67~s. However, when inspecting the light curves, one can see that significant, relatively smooth variations are present in most of them. `Hump'-like structures are especially apparent, i.e.~a smooth rise and decline lasting for $\gtrsim$1000~s, with an amplitude of up to $\sim$0.1~mag. These structures can be seen to commence at $\sim$1000~s after $T_0$ in the 11 Apr light curve, at $\sim$200~s and again at $\sim$6500~s in the 12 Apr light curve, at $\sim$2700~s on 13 Apr, and perhaps at the start of the 15 Apr observation.

In the longest CAL~83 observation, obtained on 16 December 2014, several of these humps are discernible, commencing at the following times after the observation commences ($T_0$):  $\sim$2000~s, $\sim$9500~s, $\sim$11500~s, $\sim$14000~s, and $\sim$17500~s. Similar structures with slightly smaller amplitudes and periods also seem to be present in some of the light curves. Although these rise-and-decline patterns seem to repeat themselves, they are not strictly periodic; they can be described as quasi-periodic at most. In the 16 Dec 2014 light curve, a slower rise and decline starting at $T_0\sim6000$~s and lasting for approximately 10000~s, is also evident. No deviation of more than $\sim$0.1~mag from the mean was observed within a single night. The maximum change in the mean magnitude from one night to the next is also only $\sim$0.1~mag.

Although the known orbital modulation of $\sim$1~d in the optical should be present in the SHOC light curves, the timespan of even the longest of the SHOC observations is still significantly shorter than the orbital period, therefore it is evident that variability on shorter time-scales is superimposed on the presumed underlying orbital modulation. These modulations are expected to originate in the accretion disc, which is the dominant source of optical emission in SSSs. It can be shown that the time-scales of these modulations are compatible with Keplerian periods in the intermediate to outer regions of the accretion disc, possibly of turbulent eddies that dissipate after one or two orbits. 

However, the extended rim of the accretion disc is expected to be the dominant source of (reprocessed) optical emission from the system \citep{schandl_etal1997,Meyer-Hofmeister_etal1997}. Therefore, the optical modulations may also be related to small changes in the structure of the rim itself, or perhaps the movement of blobs inside the rim.

As explained in \S\ref{sec:LIMA_DNe}, QPOs (quasi-periodic oscillations) occurring simultaneously with DNOs often exhibit the period relation $P_{\rm QPO} / P_{\rm DNO}\sim 15$. Adopting $P_{\rm DNO}\sim 67$~s leads to a period of $P_{\rm QPO} \sim 15 \times 67{\rm~s} \sim 1000$~s for a potential DNO-related QPO in the system. It is very interesting to note that this predicted time-scale is of the same order as the quasi-periodic modulations that have been observed in the SHOC light curves.

\section{Conclusions}
\label{sec:Conclusions}

The $\sim$67~s X-ray periodicity ($P_{67}$) in the supersoft X-ray source CAL~83 has been detected in all the the X-ray high state light curves obtained during the 2000--2009 time period, as well as in the X-ray low state light curve with the highest count rate. It has been shown that $P_{67}$ is of a very similar nature as the DNOs typically observed in dwarf novae during outburst. No evidence of consistent modulations at $4P_{67}$ or $8P_{67}$ that could represent associated lpDNOs or the possibly implied WD rotation, respectively, was found in the EPIC pn light curves.

Through a broad quantitative analysis, it has been shown that the well-known low-inertia magnetic accretor (LIMA) model developed by \cite{Warner1995b} and \cite{WarnerWoudt2002} for DNOs can be successfully applied to the extreme environment on the WD surface in CAL~83, in what we refer to as the `eLIMA' model. In this model, $P_{67}$ originates in a belt-like structure in the envelope at the boundary with the inner accretion disc, which is weakly coupled to the WD core by a WD magnetic field with a strength of $\sim$10$^5$~G. The belt is not in rigid corotation with the WD core, allowing it to rotate faster than the WD itself. The period variations are related to a combination of spin-up and spin-down of the belt, and the channelling of the accretion stream through different accretion arcs.

SHOC observations of CAL~83 during two different optical low states did not reveal the $\sim$67~s periodicity observed in X-rays, or any other periodic variability at high frequencies, in the optical light curves. However, the SHOC light curves of CAL~83 do exhibit significant quasi-periodic variability on longer time-scales of $\gtrsim$1000~s. If the $\sim$67~s periodicity represents the equi\-valent of a DNO, the predicted period of a DNO-related QPO in the system would be $P_{\rm QPO} \sim 15 \times 67{\rm~s} \sim 1000$~s. It is interesting that this value is of the same order as the quasi-periodic variability that has been observed.

While the feasibility of such an eLIMA model has been illustrated in this paper, a more detailed theoretical investigation and simulations of the properties of this model in supersoft X-ray binaries with their much higher accretion rates and more massive envelopes, compared to CVs, constitutes a very interesting follow-up project.

\section*{Acknowledgements}

The X-ray observations presented here were obtained with \textit{XMM-Newton}, an ESA science mission with instruments and contributions directly funded by ESA Member States and NASA. This paper also uses optical observations made at the South African Astronomical Observatory (SAAO). {\sc python} 2.7 was used for some of the data analysis in this work\footnote{python.org}. This research has made use of NASA's Astrophysics Data System.

The financial assistance of the South African Square Kilometre Array Project towards this research is hereby acknowledged. Opinions expressed and conclusions arrived at, are those of the authors and are not necessarily to be attributed to the National Research Foundation.

%%%%%%%%%%%%%%%%%%%%%%%%%%%%%%%%%%%%%%%%%%%%%%%%%%

%%%%%%%%%%%%%%%%%%%% REFERENCES %%%%%%%%%%%%%%%%%%

% The best way to enter references is to use BibTeX:

% \bibliographystyle{mnras}
% \bibliography{/home/alida/Documents/bibtex_databasis/verwysings} % if your bibtex file is called example.bib

\begin{thebibliography}{}
\makeatletter
\relax
\def\mn@urlcharsother{\let\do\@makeother \do\$\do\&\do\#\do\^\do\_\do\%\do\~}
\def\mn@doi{\begingroup\mn@urlcharsother \@ifnextchar [ {\mn@doi@}
  {\mn@doi@[]}}
\def\mn@doi@[#1]#2{\def\@tempa{#1}\ifx\@tempa\@empty \href
  {http://dx.doi.org/#2} {doi:#2}\else \href {http://dx.doi.org/#2} {#1}\fi
  \endgroup}
\def\mn@eprint#1#2{\mn@eprint@#1:#2::\@nil}
\def\mn@eprint@arXiv#1{\href {http://arxiv.org/abs/#1} {{\tt arXiv:#1}}}
\def\mn@eprint@dblp#1{\href {http://dblp.uni-trier.de/rec/bibtex/#1.xml}
  {dblp:#1}}
\def\mn@eprint@#1:#2:#3:#4\@nil{\def\@tempa {#1}\def\@tempb {#2}\def\@tempc
  {#3}\ifx \@tempc \@empty \let \@tempc \@tempb \let \@tempb \@tempa \fi \ifx
  \@tempb \@empty \def\@tempb {arXiv}\fi \@ifundefined
  {mn@eprint@\@tempb}{\@tempb:\@tempc}{\expandafter \expandafter \csname
  mn@eprint@\@tempb\endcsname \expandafter{\@tempc}}}

\bibitem[\protect\citeauthoryear{{Aly} \& {Kuijpers}}{{Aly} \&
  {Kuijpers}}{1990}]{AlyKuijpers1990}
{Aly} J.~J.,  {Kuijpers} J.,  1990, A\&A, \href
  {http://adsabs.harvard.edu/abs/1990A%26A...227..473A} {227, 473}

\bibitem[\protect\citeauthoryear{Carroll \& Ostlie}{Carroll \&
  Ostlie}{2007}]{CarrollOstlie2007}
Carroll B.,  Ostlie D.,  2007, An Introduction to Modern Astrophysics.
Pearson Addison-Wesley

\bibitem[\protect\citeauthoryear{{Cheng}, {Sion}, {Horne}, {Hubeny}, {Huang}
  \& {Vrtilek}}{{Cheng} et~al.}{1997}]{Cheng_etal1997}
{Cheng} F.~H.,  {Sion} E.~M.,  {Horne} K.,  {Hubeny} I.,  {Huang} M.,
  {Vrtilek} S.~D.,  1997, \mn@doi [AJ] {10.1086/118547}, \href
  {http://adsabs.harvard.edu/abs/1997AJ....114.1165C} {114, 1165}

\bibitem[\protect\citeauthoryear{{Coppejans} et~al.,}{{Coppejans}
  et~al.}{2013}]{Coppejans_etal2013}
{Coppejans} R.,  et~al., 2013, \mn@doi [PASP] {10.1086/672156}, \href
  {http://adsabs.harvard.edu/abs/2013PASP..125..976C} {125, 976}

\bibitem[\protect\citeauthoryear{{Elsner} \& {Lamb}}{{Elsner} \&
  {Lamb}}{1977}]{ElsnerLamb1977}
{Elsner} R.~F.,  {Lamb} F.~K.,  1977, \mn@doi [ApJ] {10.1086/155427}, \href
  {http://cdsads.u-strasbg.fr/abs/1977ApJ...215..897E} {215, 897}

\bibitem[\protect\citeauthoryear{{Eracleous} \& {Horne}}{{Eracleous} \&
  {Horne}}{1996}]{eracleous_horne1996}
{Eracleous} M.,  {Horne} K.,  1996, \mn@doi [ApJ] {10.1086/177979}, \href
  {http://adsabs.harvard.edu/abs/1996ApJ...471..427E} {471, 427}

\bibitem[\protect\citeauthoryear{{Everett} \& {Howell}}{{Everett} \&
  {Howell}}{2001}]{EverettHowell2001}
{Everett} M.~E.,  {Howell} S.~B.,  2001, \mn@doi [PASP] {10.1086/323387}, \href
  {http://adsabs.harvard.edu/abs/2001PASP..113.1428E} {113, 1428}

\bibitem[\protect\citeauthoryear{Frank, King  \& Raine}{Frank
  et~al.}{2002}]{frank_etal2002}
Frank J.,  King A.~R.,   Raine D.,  2002, {Accretion Power in Astrophysics}, 3
  edn.
Cambridge University Press, Cambridge

\bibitem[\protect\citeauthoryear{{Fujimoto}}{{Fujimoto}}{1982}]{Fujimoto1982}
{Fujimoto} M.~Y.,  1982, \mn@doi [ApJ] {10.1086/160029}, \href
  {http://adsabs.harvard.edu/abs/1982ApJ...257..752F} {257, 752}

\bibitem[\protect\citeauthoryear{{G{\"a}nsicke} \& {Beuermann}}{{G{\"a}nsicke}
  \& {Beuermann}}{1996}]{GansickeBeuermann1996}
{G{\"a}nsicke} B.~T.,  {Beuermann} K.,  1996, A\&A, \href
  {http://adsabs.harvard.edu/abs/1996A%26A...309L..47G} {309, L47}

\bibitem[\protect\citeauthoryear{{Ghosh} \& {Lamb}}{{Ghosh} \&
  {Lamb}}{1979}]{GhoshLamb1979b}
{Ghosh} P.,  {Lamb} F.~K.,  1979, \mn@doi [ApJ] {10.1086/157498}, \href
  {http://adsabs.harvard.edu/abs/1979ApJ...234..296G} {234, 296}

\bibitem[\protect\citeauthoryear{{Ghosh} \& {Lamb}}{{Ghosh} \&
  {Lamb}}{1991}]{GhoshLamb1991}
{Ghosh} P.,  {Lamb} F.~K.,  1991, {Plasma physics of accreting neutron stars}.
{Kluwer Academic Publishers}, {Dordrecht}, pp 363--444

\bibitem[\protect\citeauthoryear{{Hachisu} \& {Kato}}{{Hachisu} \&
  {Kato}}{2003}]{HachisuKato2003b}
{Hachisu} I.,  {Kato} M.,  2003, \mn@doi [ApJ] {10.1086/374968}, \href
  {http://adsabs.harvard.edu/abs/2003ApJ...590..445H} {590, 445}

\bibitem[\protect\citeauthoryear{Halliday, Resnick  \& Walker}{Halliday
  et~al.}{2005}]{Halliday_etal2005}
Halliday D.,  Resnick R.,   Walker J.,  2005, Fundamentals of physics.
Wiley international edition, Wiley

\bibitem[\protect\citeauthoryear{{Hamada} \& {Salpeter}}{{Hamada} \&
  {Salpeter}}{1961}]{hamada_salpeter1961}
{Hamada} T.,  {Salpeter} E.~E.,  1961, \mn@doi [ApJ] {10.1086/147195}, \href
  {http://adsabs.harvard.edu/abs/1961ApJ...134..683H} {134, 683}

\bibitem[\protect\citeauthoryear{Kahabka \& van~den Heuvel}{Kahabka \& van~den
  Heuvel}{2006}]{KahabkaVandenHeuvel2006}
Kahabka P.,  van~den Heuvel E. P.~J.,  2006, {Super Soft Sources}.
Cambridge University Press, New York, pp 461--474

\bibitem[\protect\citeauthoryear{{Katz}}{{Katz}}{1975}]{Katz1975}
{Katz} J.~I.,  1975, \mn@doi [ApJ] {10.1086/153788}, \href
  {http://adsabs.harvard.edu/abs/1975ApJ...200..298K} {200, 298}

\bibitem[\protect\citeauthoryear{{Lanz}, {Telis}, {Audard}, {Paerels},
  {Rasmussen}  \& {Hubeny}}{{Lanz} et~al.}{2005}]{Lanz_etal2005a}
{Lanz} T.,  {Telis} G.~A.,  {Audard} M.,  {Paerels} F.,  {Rasmussen} A.~P.,
  {Hubeny} I.,  2005, \mn@doi [ApJ] {10.1086/426382}, \href
  {http://adsabs.harvard.edu/abs/2005ApJ...619..517L} {619, 517}

\bibitem[\protect\citeauthoryear{{Long}, {Helfand}  \& {Grabelsky}}{{Long}
  et~al.}{1981}]{Long_etal1981}
{Long} K.~S.,  {Helfand} D.~J.,   {Grabelsky} D.~A.,  1981, \mn@doi [ApJ]
  {10.1086/159222}, \href {http://adsabs.harvard.edu/abs/1981ApJ...248..925L}
  {248, 925}

\bibitem[\protect\citeauthoryear{{Meintjes} \& {de Jager}}{{Meintjes} \& {de
  Jager}}{2000}]{MeintjesDeJager2000}
{Meintjes} P.~J.,  {de Jager} O.~C.,  2000, \mn@doi [MNRAS]
  {10.1046/j.1365-8711.2000.03125.x}, \href
  {http://adsabs.harvard.edu/abs/2000MNRAS.311..611M} {311, 611}

\bibitem[\protect\citeauthoryear{{Meyer-Hofmeister}, {Schandl}  \&
  {Meyer}}{{Meyer-Hofmeister} et~al.}{1997}]{Meyer-Hofmeister_etal1997}
{Meyer-Hofmeister} E.,  {Schandl} S.,   {Meyer} F.,  1997, A\&A, \href
  {http://adsabs.harvard.edu/abs/1997A%26A...321..245M} {321, 245}

\bibitem[\protect\citeauthoryear{{Ness} et~al.,}{{Ness}
  et~al.}{2015}]{Ness_etal2015}
{Ness} J.-U.,  et~al., 2015, \mn@doi [A\&A] {10.1051/0004-6361/201425178},
  \href {http://adsabs.harvard.edu/abs/2015A%26A...578A..39N} {578, A39}

\bibitem[\protect\citeauthoryear{Odendaal}{Odendaal}{2015}]{Odendaal2015}
Odendaal A.,  2015, PhD thesis, University of the Free State, Bloemfontein

\bibitem[\protect\citeauthoryear{{Odendaal} \& {Meintjes}}{{Odendaal} \&
  {Meintjes}}{2015a}]{OdendaalMeintjes2015a}
{Odendaal} A.,  {Meintjes} P.~J.,  2015a, in High Energy Astrophysics in
  Southern Africa 2014: A multi-frequency perspective of new frontiers in High
  Energy Astrophysics in Southern Africa. Mem.~S.A.It. 86(1). pp 102--107

\bibitem[\protect\citeauthoryear{{Odendaal} \& {Meintjes}}{{Odendaal} \&
  {Meintjes}}{2015b}]{OdendaalMeintjes2015b}
{Odendaal} A.,  {Meintjes} P.~J.,  2015b, Proceedings of Science,
  {PoS(FRAPWS2014)013}

\bibitem[\protect\citeauthoryear{{Odendaal}, {Meintjes}, {Charles}  \&
  {Rajoelimanana}}{{Odendaal} et~al.}{2014}]{Odendaal_etal2014a}
{Odendaal} A.,  {Meintjes} P.~J.,  {Charles} P.~A.,   {Rajoelimanana} A.~F.,
  2014, \mn@doi [MNRAS] {10.1093/mnras/stt2111}, \href
  {http://adsabs.harvard.edu/abs/2014MNRAS.437.2948O} {437, 2948}

\bibitem[\protect\citeauthoryear{{Paczy{\'n}ski}}{{Paczy{\'n}ski}}{1978}]{Paczynski1978}
{Paczy{\'n}ski} B.,  1978, in {Zytkow} A.~N.,  ed., Nonstationary Evolution of
  Close Binaries. p.~89

\bibitem[\protect\citeauthoryear{{Panei}, {Althaus}  \& {Benvenuto}}{{Panei}
  et~al.}{2000}]{Panei_etal2000}
{Panei} J.~A.,  {Althaus} L.~G.,   {Benvenuto} O.~G.,  2000, A\&A, \href
  {http://adsabs.harvard.edu/abs/2000A%26A...353..970P} {353, 970}

\bibitem[\protect\citeauthoryear{{Patterson}}{{Patterson}}{1981}]{Patterson1981}
{Patterson} J.,  1981, \mn@doi [ApJS] {10.1086/190723}, \href
  {http://adsabs.harvard.edu/abs/1981ApJS...45..517P} {45, 517}

\bibitem[\protect\citeauthoryear{{Prialnik} \& {Kovetz}}{{Prialnik} \&
  {Kovetz}}{1995}]{PrialnikKovetz1995}
{Prialnik} D.,  {Kovetz} A.,  1995, \mn@doi [ApJ] {10.1086/175741}, \href
  {http://adsabs.harvard.edu/abs/1995ApJ...445..789P} {445, 789}

\bibitem[\protect\citeauthoryear{{Rajoelimanana}, {Charles}, {Meintjes},
  {Odendaal}  \& {Udalski}}{{Rajoelimanana}
  et~al.}{2013}]{Rajoelimanana_etal2013}
{Rajoelimanana} A.~F.,  {Charles} P.~A.,  {Meintjes} P.~J.,  {Odendaal} A.,
  {Udalski} A.,  2013, \mn@doi [MNRAS] {10.1093/mnras/stt645}, \href
  {http://adsabs.harvard.edu/abs/2013MNRAS.432.2886R} {432, 2886}

\bibitem[\protect\citeauthoryear{{Reinsch}, {van Teeseling}, {King}  \&
  {Beuermann}}{{Reinsch} et~al.}{2000}]{Reinsch_etal2000}
{Reinsch} K.,  {van Teeseling} A.,  {King} A.~R.,   {Beuermann} K.,  2000,
  A\&A, \href {http://adsabs.harvard.edu/abs/2000A%26A...354L..37R} {354, L37}

\bibitem[\protect\citeauthoryear{{Schandl}, {Meyer-Hofmeister}  \&
  {Meyer}}{{Schandl} et~al.}{1997}]{schandl_etal1997}
{Schandl} S.,  {Meyer-Hofmeister} E.,   {Meyer} F.,  1997, A\&A, \href
  {http://adsabs.harvard.edu/abs/1997A%26A...318...73S} {318, 73}

\bibitem[\protect\citeauthoryear{{Seward} \& {Mitchell}}{{Seward} \&
  {Mitchell}}{1981}]{SewardMitchell1981}
{Seward} F.~D.,  {Mitchell} M.,  1981, \mn@doi [ApJ] {10.1086/158641}, \href
  {http://adsabs.harvard.edu/abs/1981ApJ...243..736S} {243, 736}

\bibitem[\protect\citeauthoryear{{Sion} \& {Urban}}{{Sion} \&
  {Urban}}{2002}]{SionUrban2002}
{Sion} E.~M.,  {Urban} J.,  2002, \mn@doi [ApJ] {10.1086/340289}, \href
  {http://adsabs.harvard.edu/abs/2002ApJ...572..456S} {572, 456}

\bibitem[\protect\citeauthoryear{{Sion}, {Cheng}, {Huang}, {Hubeny}  \&
  {Szkody}}{{Sion} et~al.}{1996}]{Sion_etal1996}
{Sion} E.~M.,  {Cheng} F.-H.,  {Huang} M.,  {Hubeny} I.,   {Szkody} P.,  1996,
  \mn@doi [ApJ] {10.1086/310318}, \href
  {http://adsabs.harvard.edu/abs/1996ApJ...471L..41S} {471, L41}

\bibitem[\protect\citeauthoryear{{Southwell}, {Livio}, {Charles}, {O'Donoghue}
  \& {Sutherland}}{{Southwell} et~al.}{1996}]{Southwell_etal1996a}
{Southwell} K.~A.,  {Livio} M.,  {Charles} P.~A.,  {O'Donoghue} D.,
  {Sutherland} W.~J.,  1996, \mn@doi [ApJ] {10.1086/177931}, \href
  {http://adsabs.harvard.edu/abs/1996ApJ...470.1065S} {470, 1065}

\bibitem[\protect\citeauthoryear{{Str{\"u}der} et~al.,}{{Str{\"u}der}
  et~al.}{2001}]{struder_etal2001}
{Str{\"u}der} L.,  et~al., 2001, \mn@doi [A\&A] {10.1051/0004-6361:20000066},
  \href {http://adsabs.harvard.edu/abs/2001A%26A...365L..18S} {365, L18}

\bibitem[\protect\citeauthoryear{{Szkody}, {Hoard}, {Sion}, {Howell}, {Cheng}
  \& {Sparks}}{{Szkody} et~al.}{1998}]{Szkody_etal1998}
{Szkody} P.,  {Hoard} D.~W.,  {Sion} E.~M.,  {Howell} S.~B.,  {Cheng} F.~H.,
  {Sparks} W.~M.,  1998, \mn@doi [ApJ] {10.1086/305506}, \href
  {http://adsabs.harvard.edu/abs/1998ApJ...497..928S} {497, 928}

\bibitem[\protect\citeauthoryear{{Townsley} \& {Bildsten}}{{Townsley} \&
  {Bildsten}}{2004}]{TownsleyBildsten2004}
{Townsley} D.~M.,  {Bildsten} L.,  2004, \mn@doi [ApJ] {10.1086/379647}, \href
  {http://adsabs.harvard.edu/abs/2004ApJ...600..390T} {600, 390}

\bibitem[\protect\citeauthoryear{{Tr{\"u}mper} et~al.,}{{Tr{\"u}mper}
  et~al.}{1991}]{Trumper_etal1991}
{Tr{\"u}mper} J.,  et~al., 1991, \mn@doi [Nature] {10.1038/349579a0}, \href
  {http://adsabs.harvard.edu/abs/1991Natur.349..579T} {349, 579}

\bibitem[\protect\citeauthoryear{{Turner} et~al.,}{{Turner}
  et~al.}{2001}]{turner_etal2001}
{Turner} M.~J.~L.,  et~al., 2001, \mn@doi [A\&A] {10.1051/0004-6361:20000087},
  \href {http://adsabs.harvard.edu/abs/2001A%26A...365L..27T} {365, L27}

\bibitem[\protect\citeauthoryear{{Van den Heuvel}, {Bhattacharya}, {Nomoto}  \&
  {Rappaport}}{{Van den Heuvel} et~al.}{1992}]{VandenHeuvel_etal1992}
{Van den Heuvel} E.~P.~J.,  {Bhattacharya} D.,  {Nomoto} K.,   {Rappaport}
  S.~A.,  1992, A\&A, \href
  {http://adsabs.harvard.edu/abs/1992A%26A...262...97V} {262, 97}

\bibitem[\protect\citeauthoryear{Warner}{Warner}{1995a}]{Warner1995a}
Warner B.,  1995a, {Cataclysmic Variable Stars}, 2 edn.
No.~28 in Cambridge Astrophysics Series, Cambridge University Press, Cambridge

\bibitem[\protect\citeauthoryear{{Warner}}{{Warner}}{1995b}]{Warner1995b}
{Warner} B.,  1995b, in {Buckley} D.~A.~H.,  {Warner} B.,  eds,  Astronomical
  Society of the Pacific Conference Series Vol. 85, Magnetic Cataclysmic
  Variables. p.~343

\bibitem[\protect\citeauthoryear{{Warner}}{{Warner}}{2004}]{warner2004}
{Warner} B.,  2004, \mn@doi [PASP] {10.1086/381742}, \href
  {http://adsabs.harvard.edu/abs/2004PASP..116..115W} {116, 115}

\bibitem[\protect\citeauthoryear{{Warner} \& {Woudt}}{{Warner} \&
  {Woudt}}{2002}]{WarnerWoudt2002}
{Warner} B.,  {Woudt} P.~A.,  2002, \mn@doi [MNRAS]
  {10.1046/j.1365-8711.2002.05596.x}, \href
  {http://adsabs.harvard.edu/abs/2002MNRAS.335...84W} {335, 84}

\bibitem[\protect\citeauthoryear{{Warner} \& {Woudt}}{{Warner} \&
  {Woudt}}{2008}]{WarnerWoudt2008}
{Warner} B.,  {Woudt} P.~A.,  2008, in {Axelsson} M.,  ed.,  American Institute
  of Physics Conference Series Vol. 1054, American Institute of Physics
  Conference Series. pp 101--110 (\mn@eprint {arXiv} {0806.1317}),
  \mn@doi{10.1063/1.3002491}

\bibitem[\protect\citeauthoryear{{Warner}, {Woudt}  \& {Pretorius}}{{Warner}
  et~al.}{2003}]{Warner_etal2003}
{Warner} B.,  {Woudt} P.~A.,   {Pretorius} M.~L.,  2003, \mn@doi [MNRAS]
  {10.1046/j.1365-8711.2003.06905.x}, \href
  {http://adsabs.harvard.edu/abs/2003MNRAS.344.1193W} {344, 1193}

\bibitem[\protect\citeauthoryear{{XMM-Newton Science Operations Centre
  Team}}{{XMM-Newton Science Operations Centre Team}}{2014}]{sas_manual2014}
{XMM-Newton Science Operations Centre Team} 2014, {Users Guide to the
  XMM-Newton Science Analysis System Issue 10.5}, Available at:
  \url{http://xmm.esac.esa.int/external/xmm_user_support/documentation/sas_usg/USG.pdf}
  (accessed 01-02-2015).

\makeatother
\end{thebibliography}

%%%%%%%%%%%%%%%%%%%%%%%%%%%%%%%%%%%%%%%%%%%%%%%%%%

%%%%%%%%%%%%%%%%% APPENDICES %%%%%%%%%%%%%%%%%%%%%

%%%%%%%%%%%%%%%%%%%%%%%%%%%%%%%%%%%%%%%%%%%%%%%%%%

% Don't change these lines
\bsp	% typesetting comment
\label{lastpage}
\end{document}